\documentclass{amsart} 
\usepackage{mathrsfs,amssymb,amsmath,amsxtra,amsfonts}
\usepackage{graphicx,color}
\usepackage{verbatim}
\usepackage[colorlinks]{hyperref}

\begin{document}
\newcommand{\A  }{\mbox{$\textrm{Aff}(3)$}}  
\renewcommand{\a}{\alpha}
\renewcommand{\b}{\beta}
\newcommand{\x}{\mbox{$\mathfrak{x}$}}
\renewcommand{\t}{\mbox{$\mathfrak{t}$}}
\newcommand{\R}{\mbox{$\mathbb{R}$}}
\newcommand{\C}{\mbox{$\mathbb{C}$}}
\newcommand{\pd}{\mbox{$\partial$}} 
\providecommand{\abs}[1]{\lvert#1\rvert}

\newtheorem{theorem}{Theorem}
\newtheorem{axiom}{Axiom}
\newtheorem{definition}{Definition}
\newtheorem{thesis}{Thesis}
\newtheorem{corollary}{Corollary}
\author{I. Schmelzer}
\thanks{Berlin, Germany}
\email{ilja.schmelzer@gmail.com}%
\urladdr{ilja-schmelzer.de}

\title[Generalization of Lorentz ether to gravity]{A generalization of the Lorentz ether to gravity with
general-relativistic limit} \sloppypar\sloppy

\begin{abstract}
Does relativistic gravity provide arguments against the existence of a preferred frame? Our answer is negative. We define a viable theory of gravity with preferred frame. In this theory, the EEP holds exactly, and the Einstein equations of GR limit are obtained in a natural limit. Despite some remarkable differences (stable ``frozen stars'' instead of black holes, a ``big bounce'' instead of the big bang, exclusion of nontrivial topologies and closed causal loops, and a preference for a flat universe) the theory is viable. 

The equations of the theory are derived from simple axioms about some fundamental condensed matter (the generalized Lorentz ether), so that, in particular, the EEP is not postulated but derived. 

The theory is compatible with the condensed matter interpretation for the fermions and gauge fields of the standard model.

\end{abstract}

\maketitle

\section{Motivation}\label{motivation}

We propose here to revive the concept of an ether -- a concept which has been rejected more than a century ago. This is an extraordinary proposal which requires extraordinary motivation.

In my opinion, the most important motivation comes from a condensed matter interpretation of the fermions and gauge fields of the standard model of particle physics proposed in \cite{clm}. A key of this interpretation is a three-dimensional geometric interpretation of the fermions of the standard model as sections of the bundle $\A\otimes\C\otimes\Lambda(\R^3)$, where each electroweak doublet is defined by a section of the complexified exteriour bundle $\C\otimes\Lambda(\R^3)$, and the geometric Dirac operator on this bundle is used to define the Dirac equation. In this interpretation, geometric rotations are identified with a combination of spinor rotations and isospin rotations. This interpretation cannot be generalized to a four-dimensional variant -- the four-dimensional Dirac-K\"ahler equation gives four Dirac fermions instead of two. So the viability of this geometric model depends on the viability of a preferred frame in relativistic gravity. 

But there is also a sufficiently large number of other motivations. The existence of a viable ether theory in the domain of relativistic gravity would be interesting already in itself, even if only for playing devil's advocate. It would force us to reevaluate the arguments against the ether. In particular, the strongest empirical argument against the ether -- non-existence of a viable ether theory of gravity -- would become invalid. Some other popular arguments against the old Lorentz ether will shown to be invalid for the theory proposed here: Relativistic symmetry is no longer ad hoc --- in our theory, we derive the Einstein equivalence principle from independent simple axioms.  Then, the violation of the ``action equals reaction'' principle by the static, incompressible Lorentz ether no longer exists in the theory presented here, which has a classical Lagrange formalism.

A strong motivation comes from the foundations of quantum theory. During the last years, the de Broglie-Bohm pilot wave interpretation has become increasingly popular. It is a realistic, deterministic, and causal interpretation of quantum theory. The main argument against this interpretation is that it needs, in the relativistic domain, a hidden preferred frame. While the introduction of such a hidden preferred frame is not problematic in the special-relativistic domain, the situation is different in the domain of relativistic gravity, where, if GR is true, we can even have manifolds with non-trivial topology and other artefacts like closed causal loops, which would forbid the introduction of a preferred frame. Thus, pilot wave interpretations seem incompatible with GR. Therefore, the viability of the pilot wave interpretation depends on the possibility to modify relativistic gravity in such a way that it becomes compatible with a preferred frame.

This is not only a particular problem of a particular interpretation of quantum theory. The violation of Bell's inequality proves that any realistic interpretation of quantum theory needs a preferred frame. Thus, if a preferred frame would be incompatible with relativistic gravity, we would have to give up not only the pilot wave interpretation, but even realism --- a very fundamental scientific principle which can be even interpreted as an implicit part of the scientific method: Giving up realism means giving up the search for realistic explanations of observable phenomena. Before giving up such a fundamental scientific principle, the alternatives should be evaluated carefully. 

This problem is often represented in a wrong way, as realism being in conflict with quantum theory. But pilot wave interpretations prove the compatibility of realism with quantum theory. The conflict is one between realism and relativity, in particular GR, which does not allow the introduction of a preferred frame. Therefore, the viability of realism --- one of the foundations of science --- depends on the possibility to develop a theory of gravity with preferred frame too.

Then, the problem of quantization of gravity gives another independent motivation for an ether interpretation. While this problem is claimed to be solved in string theory, at the current moment the situation in string theory does not look very promising \cite{Woit}, and it seems quite possible that string theory fails to give a reasonable theory of everything. In this case, the problem of quantization of gravity would be open again. Now, we know very well how to quantize condensed matter theories. Therefore, if we find a way to interpret relativistic gravity as a condensed matter theory, we can apply the large amount of knowledge about quantum condensed matter theories to quantize gravity.

Thus, there are very different motivations for a revival of the ether. The main questions raised by these different motivations are the following two: First, is the existence of a hidden preferred frame compatible with relativistic gravity? Second, is it possible to find some condensed matter interpretation for the gravitational field? The theory we present here gives a positive answer to above questions. Moreover, the condensed matter interpretation we propose here for gravity seems compatible with the ether interpretation of the standard model proposed in \cite{clm}.

In an appendix of \cite{clm}, a short introduction into this theory has been already presented. But this introduction was much too short, and leaves many questions unanswered. The aim of this paper is to give a more detailed introduction into this theory. Moreover, we want to present some consequences (in particular a big bounce instead of the big bang and stable gravastars instead of black holes) and discuss the connection of our theory with other theories, in particular two theories which share with our theory the same Lagrangian, namely, general relativity with four scalar ``clock fields'' \cite{Kuchar}, and the relativistic theory of gravity \cite{Logunov1}. Last but not least, we have to discuss various possible objections.

The main part of this paper contains all what is necessary to define the theory. Most of the other points we have considered in various appendices. Some of these appendices, especially the parts about analog models, quantization, realism and pilot wave theory, and the predictions, have to be considered in separate papers in more detail. While we focus our attention on the presentation of an ether theory of gravity, the main technical results -- the derivation of a metric theory with GR limit from condensed-matter-like axioms -- may be, possibly, applied also in a completely different domain: Search for classical condensed matter theories which fit into this set of axioms as analog models of general relativity. This possibility we discuss in app. \ref{appCondensedMatter}.

\section{Introduction}

Thus, the aim of this paper is to propose a generalization of the classical Lorentz ether to gravity. It is a theory on Euclidean space and absolute time $\R^3\otimes \R$, with preferred coordinates. They will be denoted $\x^i,\t$ in contexts where the preferred frame (absolute time) is important, and $\x^\a$ (with $\t=\x^0$) for the preferred space-time coordinates. We also consider general systems of coordinates, which will be denoted with $x^\mu$.  We use the corresponding indices to denote partial derivatives: latin indices $\pd_i$ for derivatives $\frac{\pd}{\pd \x^i}$ in the preferred spatial coordinates, $\pd_0$ for $\frac{\pd}{\pd \t}= \frac{\pd}{\pd \x^0}$ in the preferred time, lower greek indices $\pd_\a$ for $\frac{\pd}{\pd \x^\a}$ in preferred coordinates, and larger greek indices $\pd_\mu$ for $\frac{\pd}{\pd x^\mu}$ in general coordinates. Similar conventions hold for other tensor indices, so that $g^{ij}$ denotes only the $3\times 3$ matrix of spatial components of $g^{\a\b}(\x^\a)$ in preferred coordinates.

The theory, which we have named ``General Lorentz Ether Theory'' (GLET), is a metric theory of gravity, with a Lagrangian, which has the GR Lagrangian as a limit $\Xi,\Upsilon\to 0$:
\begin{equation}\label{L}
L = L_{GR} + L_{matter}(g_{\mu\nu},\varphi^m) 
  - (8\pi G)^{-1} (\Upsilon g^{00}-\Xi \delta_{ij}g^{ij})\sqrt{-g}.
\end{equation}
This Lagrangian is not covariant, thus, a preferred system of coordinates $\x^\a$ and in particular a preferred time $\t=\x^0$ are presupposed. 

On the other hand, the theory may be interpreted as a classical ether theory, that means, a classical condensed matter theory in a Newtonian framework $\mathbb{R}^3\otimes\mathbb{R}$: We assume classical condensed matter variables (density $\rho(\x,\t)$, velocity $v^i(\x,\t)$, pressure tensor $p^{ij}(\x,\t)$), which follow classical conservation laws (Euler and continuity equations):
\begin{subequations}\label{classical}
\begin{align}
\partial_0 \rho + \partial_i (\rho v^i) &= 0 \\
\partial_0 (\rho v^j) + \partial_i(\rho v^i v^j + p^{ij}) &= 0,
\end{align}
\end{subequations}
together with a Lagrange formalism and a connection between them as required by Noether's theorem.  The only ``strange'' assumption is that the pressure tensor $p^{ij}$ should be negative definite.

The transformation of this theory into a metric theory of gravity is defined in the preferred coordinates $\x^\a$ by the following simple definition of the effective gravitational field $g_{\a\b}(\x^\a)$:
\begin{subequations}\label{gdef}
\begin{align}
\hat{g}^{00} = g^{00}\sqrt{-g} &= \rho, \\
\hat{g}^{i0} = g^{0i}\sqrt{-g} &= \rho v^i, \\
\hat{g}^{ij} = g^{ij}\sqrt{-g} &= \rho v^i v^j + p^{ij}.
\end{align}
\end{subequations}
Matter fields $\varphi^m$ have to be identified with other degrees of freedom (material properties) of the ether.  The axioms of the theory, as far as considered here, do not specify the material properties of the ether. Therefore, they do not fix the matter degrees of freedom and the matter Lagrangian. There may be different ether models, which lead to theories with different matter fields. For a particular proposal for the fermions and gauge fields of the standard model of particle physics see \cite{clm}. Nonetheless, the general axioms considered here are not only sufficient to derive the gravitational part of the general Lagrangian \eqref{L}, but also allow to derive the Einstein equivalence principle (EEP) for the matter fields. Moreover, we obtain well-defined conservation laws with local energy and momentum densities for the gravitational field.

\section{Conservation Laws as Euler-Lagrange Equations for Preferred Coordinates}

Below we consider classical condensed matter theories with Lagrange formalism and classical conservation laws as known from condensed matter theory (continuity and Euler equations).  We know from Noether's theorem that translational invariance of the Lagrange density leads to conservation laws.  It seems natural to identify the conservation laws obtained via Noether's theorem from the Lagrange formalism with the conservation laws as known them from condensed matter theory.  The problem with this identification is that the conservation laws are not uniquely defined.  If a tensor $T^{\alpha\beta}$ fulfills $\partial_\alpha T^{\alpha\beta}=0$, then the tensor $T^{\alpha\beta}+\partial_\gamma \psi^{\alpha\beta\gamma}$ fulfills the same equation if $\psi$ is an arbitrary tensor which fulfills the condition $\psi^{\alpha\beta\gamma}=-\psi^{\gamma\beta\alpha}$.

We solve this problem by proposing a variant of Noether's theorem which seems privileged because it is especially simple, natural and beautiful. Then this especially beautiful variant we identify with the known, also especially simple, classical conservation laws from condensed matter theory --- the continuity and Euler equations.

The theory we consider here depends on a preferred frame, or, in other words, on preferred coordinates. These preferred coordinates $\x^\a$ are, themself, functions on the manifold $\mathbb{R}^3\otimes\mathbb{R}$ of the theory.
Now, once the Lagrangian depends on the preferred coordinates, it seems natural to ask if we can obtain Euler-Lagrange equations for them. This is, indeed, possible. There are two points which have to be taken into account. First, the technical one, that not all possible variations $\x^\a(x) \to \x^\a(x) + \delta \x^\a(x)$ are allowed --- the result of the variation should be, again, a system of coordinates. But, as we show in appendix \ref{appCoordinates}, this restriction does not prevent the derivation of the Euler-Langrange equations. The other, conceptually more important one is that if a classical theory is defined in a preferred frame, it may be simply not defined in other, non-preferred frames. If no such definition exists, there is also no dependence of the Lagrangian on the preferred coordinates.

Partially, this problem can be solved. Let's consider the meaning of the terms in the Euler-Langrange equation
\begin{equation}
 \frac{\delta S}{\delta \x^\a} = \frac{\pd L}{\pd \x^\a} - \pd_\mu\left( \frac{\pd L}{\pd \x^\a_{,\mu}} - \cdots\right).
\end{equation} 
Assume, for example, that there is some vector field $u^\mu(x)$ among the variables, and assume that the Lagrangian, written in the preferred coordinates, depends on $u^0(\x)$. The expression $u^0(\x)$ clearly depends on the preferred coordinates. But what is, in this case, $\frac{\pd u^0}{\pd \x^\a}$ or $\frac{\pd u^0}{\pd \x^\a_{,\mu}}$? This question can, fortunately, be solved in a simple way: Namely, we have $u^0(\x) = u^\a(\x)\x^0_{,\a}(\x)$. In the last expression, we have already an explicit dependence on $\x^0_{,\mu}$. Moreover, there is already no other dependence on the preferred coordinates --- the expression is, if we forget for a moment about the special character of $\x^0$, a formally covariant expression, we can replace the coordinates $\x$ by other coordinates $x$ without modifying their form: $u^\a(\x)\x^0_{,\a}(\x)=u^\mu(x)\x^0_{,\mu}(x)$.

It is worth to note at this place that the preferred coordinates do not define a single vector function, as suggested by their upper index $\x^\a(x)$, but, instead, are four independent scalar functions on the manifold: Indeed, if we consider them as functions $\x^\a(x)$ on the manifold, they transform like scalar fields if we consider transformations of coordinates $x \to x'=x'(x)$: Their transformation law is simply $\x^\a(x'(x)) = \x^\a(x)$. If they would transform as a vector field, the transformation law would be $\x^{\a'}(x'(x)) = \frac{\pd x^{\a'}}{\pd x^\a}\x^\a(x)$, which would be wrong.

The form of presentation similar to $u^\mu(x)\x^0_{,\mu}(x)$ we name \emph{explicit dependence on the preferred coordinates}. Formally, this can be defined in the following way:

\begin{definition} The dependence of an expression $F(\phi^k,\phi_{,\mu}^k,\ldots,\x^\alpha,\x^\alpha_{,\mu},\ldots)$ on the preferred coordinates $\x^\alpha$ is \emph{explicit} if the expression $F(\phi^k,\phi_{,i}^k,\ldots,U^\alpha,U^\alpha_{,\mu},\ldots)$, where the coordinates $\x^\alpha(x)$ have been replaced by four independent scalar fields $U^\alpha(x)$, is covariant.\footnote{This includes the case of covariance -- no dependence on the preferred coordinates at all.}
\end{definition}

Indeed, if the second expression is covariant, no implicit dependence on the preferred coordinates $\x^\a$ has been left.  Now, it may be asked if every dependence on preferred coordinates can be made explicit.  In this context, we propose the following

\begin{thesis}\label{thesis:explicit}
For every physically interesting function $F(\phi^k,\phi_{,\mu}^k,\ldots)$ there exists a function $\hat{F}(\phi^k,\phi_{,\mu}^k,\ldots,\x^\alpha,\x^\alpha_{,\mu},\ldots) = F(\phi^k,\phi_{,\mu}^k,\ldots)$ so that any dependence on the preferred coordinates $\x^\alpha$ is explicit.
\end{thesis}

Note that the thesis is not obvious. The definition of explicit dependence already requires that we know how to transform the other fields $\phi^k$. For a non-covariant theory this is possibly not yet defined (see app. \ref{appGeneral}). In this sense, this thesis may be considered as analogous to Kretschmann's thesis \cite{Kretschmann} that every physical theory allows a covariant formulation. But we do not have to assume (and do not assume in the following) that this thesis holds.  If it holds, the assumption of axiom \ref{aLagrange} is less restrictive.

But one has to note that a unique transformation of a non-covariant expression is possible only if the transformation law of the expression is well-defined. A scalar function is covariant in itself, a vector component $u^0(x)$ is equivalent to $u^\mu(x)\x^0_{,\mu}(x)$, a tensor component $u^{00}(x)$ equivalent to $u^{\mu\nu}(x)\x^0_{,\mu}(x)\x^0_{,\nu}(x)$. But what to do if we have some entity like, for example, a density $\rho(x)$?  Is it a scalar? Or is it the component $p^0$ of a vector $p^0=\rho(x), p^i=\rho(x) v^i(x)$? Or is it a component $t^{00}(x)=\rho(x)$ of the energy-momentum tensor? These questions have to be answered before one can define an explicit dependence of a theory on the preferred coordinates.

\subsection{Euler-Lagrange Equations for Preferred Coordinates and Noether's Theorem}

Now, if the dependence of the Lagrangian on the preferred coordinates is explicit, we can vary not only over the fields, but also over the preferred coordinates --- all terms in the Euler-Lagrange equations have now a well-defined meaning, the same meaning as for usual fields. Thus, as a consequence we obtain Euler-Lagrange equations for the preferred coordinates $\x^\alpha$:

\begin{theorem}\label{Euler-Lagrange}
If the Lagrange density $L(\phi^k,\phi_{,\mu}^k,\ldots,\x^\alpha,\x^\alpha_{,\mu},\ldots)$ explicitly depends on the preferred coordinates $\x^\a$, then for the extremum of $S = \int L$ the following Euler-Lagrange equation holds:

\[ \frac{\delta S}{\delta \x^\a} = \frac{\partial L}{\partial \x^\a}
- \partial_\mu \frac{\partial L}{\partial \x^\a_{,\mu}}
+ \partial_\mu\partial_\nu \frac{\partial L}{\partial \x^\a_{,\mu\nu}}
-\cdots = 0. 
\]
\end{theorem}

For the proof, see appendix \ref{appCoordinates}.  This does not mean that the special geometric nature of the $\x^\alpha$ is not important. We see in appendix \ref{appGRCF} that a theory with ``the same'' Lagrangian but usual scalar fields $U^\alpha(x)$ instead of coordinates $\x^\alpha(x)$ is a completely different physical theory.

We know from Noether's theorem that symmetries of the Lagrangian lead to conservation laws.  Especially, conservation of energy and momentum follows from translational symmetry in the preferred coordinates. Now, in the case of explicit dependence on the preferred coordinates $\x^\alpha$, translational symmetry $\x^\alpha\to \x^\alpha+c^\alpha$ has an especially simple form.  We obtain the conservation laws automatically as Euler-Lagrange equations for the coordinates! Indeed, translational symmetry immediately gives the Lagrangian does
not depend on the $\x^\alpha$ them-self, only on their partial derivatives $\x^\a_{,\mu},\x^\a_{,\mu\nu}\ldots$.  Therefore the Euler-Lagrange equations for the $\x^\alpha$ \emph{already have the form of conservation laws}.

\begin{theorem}\label{Noether1}

If a Lagrangian with explicit dependence on the preferred coordinates has translational symmetry $\x^\alpha \to \x^\alpha + c^\alpha$, then the Euler-Lagrange equations for the preferred coordinates $\x^\alpha$ become conservation laws:
\begin{equation}
\frac{\delta S}{\delta \x^\alpha} = \partial_\mu {T^\mu_\alpha}
\end{equation}
with
\begin{equation}
T^\mu_\alpha = -\frac{\partial L}{\partial \x^\a_{,\mu}}
+ \partial_\nu \frac{\partial L}{\partial \x^\a_{,\mu\nu}}
-\cdots.
\end{equation} 
\end{theorem}

We know that in agreement with Noether's second theorem these conservation laws disappear if we have general covariance.  Again, in our formulation we do not have to do much to see this -- if there is no dependence on the $\x^\alpha$, the Euler-Lagrange equation for the $\x^\alpha$, which are the conservation laws, disappear automatically:

\begin{theorem}\label{Noether2}
If L is covariant, then
\begin{equation} \label{eqNoether2}
\frac{\delta S}{\delta \x^\alpha} \equiv 0 
\end{equation}
\end{theorem}

In above cases, the proof is an obvious consequence of theorem \ref{Euler-Lagrange}.  Of course, there is not much to wonder about, we have simply used a set of variables $\x^\alpha$ appropriate for the symmetries we have considered -- translational symmetry.  

Based on these theorems we identify the Euler-Lagrange equation $\frac{\delta S}{\delta \x^0}$ with the {\em energy conservation law\/} and $\frac{\delta S}{\delta \x^i}$ with the {\em momentum conservation law\/}. Note also that the most general GR Lagrangian is given by the conditions (\ref{eqNoether2}).

It seems worth to note that until now we have only considered a quite general formalism which is in no way restricted to the application in condensed matter theory below.

\section{The ADM decomposition}\label{ADM}

The definition \eqref{gdef} of $g_{\a\b}(\x)$ in terms of three-dimensional objects $\rho, v^i, p^{ij}$ is, in fact, a variant of the ADM decomposition \cite{ADM}. The ADM decomposition splits the four-metric into a scalar field, a vector field, and a three-metric. What is added here is the particular identification of these three-dimensional objects with objects of condensed matter theory. 

We have to check, of course, that the transformation laws for the objects of above sides of \eqref{gdef} are compatible with each other.  Here one may doubt that everything is correct given that the density $\rho(\x)$ is a three-density, but $g^{00}\sqrt{-g}$ a component of a four-density. 

But the transformation laws for condensed matter fields are well-defined only for transformations which leave absolute time unchanged. So, the question of correctness of the transformation laws makes sense only for such transformations. But in this case everything is fine, and spatial densities transform in the same way as four-densities. 

It is important to emphasize here the decisive role of the preferred coordinates $\x^i,\t$. They define a preferred one-form $d\t$ as well as a preferred vector field $\pd_0$ which distinguishs absolute rest. These special objects may be used to define four-dimensional objects out of three-dimensional ones. In particular, a three-vector field $v^i\pd_i$ can be identified with $v^\a\pd_a=\pd_0+v^i\pd_i$, which is already a legitimate four-vector field. Similarly, a three-density $\rho(\x,\t) d\x^1\land d\x^2 \land d\x^3$ can be identified with the four-density $\rho(\x,\t) d\t\land d\x^1\land d\x^2 \land d\x^3$, which is a legitimate four-density. 

With these identifications, the condensed matter objects on the right-hand-side of \eqref{gdef} can be identified with a densitised energy-momentum tensor field $t^{\a\b}(\x)$ defined by
 
\begin{subequations}\label{tdef}
\begin{align}
t^{00} &= \rho, \\
t^{0i} &= \rho v^i, \\
t^{ij} &= \rho v^i v^j + p^{ij}.
\end{align}
\end{subequations}

For this tensor field, the transformation law is defined already for arbitrary coordinate transformations, and $t^{\mu\nu}$ transforms in the same way as $g^{\mu\nu}\sqrt{-g}$, so that \eqref{gdef} can be extended to arbitrary systems of coordinates as $t^{\mu\nu}=g^{\mu\nu}\sqrt{-g}$. 

\section{Axioms for General Lorentz Ether Theory}

Let's define now our set of axioms for the class of condensed-matter-like theories we call ``Lorentz ether theories''.  It looks quite natural, nonetheless, it should be noted that it is in some essential points an unorthodox formalism for condensed matter theories.  Especially it remains open if there exist classical condensed matter theories which fit into this set of axioms.

\begin{axiom}[independent variables] \label{afirst} The independent variables of the theory are the following fields defined on a Newtonian framework ${\mathbb R^3\otimes R}$ with preferred coordinates $\x^i,\t$: a positive density $\rho(\x^i,\t)$, a velocity $v^i(\x^i,\t)$, a symmetric pressure tensor $p^{ij}(\x^i,\t)$.

Moreover, there is an unspecified number of ``inner degrees of freedom'' (material properties) $\varphi^m(\x^i,\t)$ with well-defined transformation rules for arbitrary coordinate transformations.
\end{axiom}

That we use a pressure tensor $p^{ij}$ instead of a scalar $p$ is a quite natural and known generalization.  Instead, the use of $p^{ij}$ as an independent variable is non-standard.  Usually pressure is defined as an algebraic ``material function'' of other independent variables.

\begin{axiom}[Lagrange formalism] \label{aLagrange}
There exists a Lagrange formalism with explicit dependence of the Lagrange density on the preferred coordinates
\begin{equation*}
L = L(t^{\mu\nu},t^{\mu\nu}_{,\kappa},\ldots, \varphi^m,\varphi^m_{,\kappa},\ldots, \x^\alpha,\x^\alpha_{,\kappa},\ldots)
\end{equation*}
\end{axiom}

The definition of explicit dependence requires, of course, a specification of the general transformation law for the variables. Here it is, in agreement with the considerations of sec. \ref{ADM}, presupposed that this is the transformation law for a densities energy-momentum tensor $t^{\mu\nu}$ defined in the preferred coordinates by \eqref{tdef}. 

The existence of a Lagrange formalism is certainly a non-trivial physical restriction (it requires an ``action equals reaction'' symmetry) but is quite common in condensed matter theory as far as reversible, elastic phenomena are considered.  Instead, the requirement of explicit dependence on the $\x^\a$ is non-standard. In appendix \ref{appGeneral} we argue that this additional requirement is not really a restriction: Every physical theory with Lagrange formalism can be transformed into this form.  But this remains a general thesis which cannot be proven in a strong way, only supported by showing how this can be done in particular cases.

\begin{axiom}[energy conservation law] \label{acontinuity}
The conservation law related by theorem \ref{Noether1} with translational symmetry in time $\x^0\to \x^0+c^0$ is proportional to the continuity equation:
\begin{equation}
\frac{\delta S}{\delta \x^0} \sim \partial_0 \rho + \partial_i (\rho v^i).
\end{equation}
\end{axiom}

\begin{axiom}[momentum conservation law] \label{aEuler} 
The conservation laws related by theorem \ref{Noether1} with translational symmetry in space $\x^i\to \x^i+c^i$ are proportional to the Euler equation:
\begin{equation}
\frac{\delta S}{\delta \x^j} \sim \partial_0 (\rho v^j) + \partial_i(\rho v^i v^j+p^{ij}).
\end{equation}
\end{axiom}

These axioms also look quite natural and are well motivated by our version of Noether's theorem.  But it should be noted that this is not the only possibility to connect continuity and Euler equations with the Lagrange formalism.  The energy-momentum tensor is not uniquely defined.  The main purpose of the introduction of our ``explicit dependence'' formalism was to motivate this particular choice as the most natural one.\footnote{If a tensor $T^{\alpha\beta}$ fulfills $\partial_\alpha T^{\alpha\beta}=0$, then the tensor $T^{\alpha\beta}+\partial_\gamma \psi^{\alpha\beta\gamma}$ fulfills the same equation if $\psi$ is an arbitrary tensor which fulfills the condition $\psi^{\alpha\beta\gamma}=-\psi^{\gamma\beta\alpha}$.}  Thus, by our choice of axioms \ref{acontinuity}, \ref{aEuler} we fix a particular, even if especially simple, relation between Lagrange formalism and continuity and Euler equations.

Despite the non-standard technical features of these axioms, none of them looks like being made up to derive the results below.  And we also have not fixed anything about the inner structure of the medium: The inner degrees of freedom (material properties) and the material equations are not specified.  From point of view of physics, the following non-trivial restrictions have been made:

\begin{itemize}

\item The existence of a Lagrange formalism requires a certain symmetry -- the equations are self-adjoint (the principle ``action equals reaction'').

\item This medium is conserved (continuity equation).

\item There is only one, universal medium.  There are no external forces and there is no interaction terms with other types of matter (as follows from the Euler equation).

\end{itemize}

Last but not least, there is one additional axiom which sounds quite strange:

\begin{axiom}[negative pressure] \label{alast}\label{apressure}
The pressure tensor $p^{ij}(\x^i,\t)$ is negative definite
\end{axiom}

Instead, for most usual matter pressure is positive.  On the other hand, the Euler equation defines pressure only modulo a constant, and if we add a large enough negative constant tensor $-C\delta^{ij}$ we can make the pressure tensor negative definite.  The physical meaning of this axiom is not clear.

These are all axioms we need.  Of course, these few axioms do not define the theory completely.  The material properties are not defined.  Thus, our axioms define a whole class of theories instead of a particular theory.  Every medium which meets these axioms we name ``Lorentz ether''.

\section{The Derivation of the General Lagrangian}

The technical key result of the paper is the following derivation:

\begin{theorem}\label{main}
Assume we have a theory which fulfills the axioms
\ref{afirst}-\ref{alast}.

Then there exists an ``effective metric'' $g_{\mu\nu}(x)$ of signature $(1,3)$, which is uniquely defined by $\left\{\rho, v^i, p^{ij}\right\}$, and some constants $\Xi,\Upsilon$ so that for every solution $\left\{\rho(\x,\t),v^i(\x,\t),p^{ij}(\x,\t),\varphi^m(\x,\t)\right\}$ the related fields $g_{\mu\nu}(x),\varphi^m(x)$ together with the preferred coordinates $\x^\alpha(x)$ are also solutions of the Euler-Lagrange equations of the Lagrangian {\rm }

\begin{equation}
L = -(8\pi G)^{-1}\gamma_{\alpha\beta}\x^\alpha_{,\mu}\x^\beta_{,\nu}
        g^{\mu\nu}\sqrt{-g} + L_{GR}(g_{\mu\nu}) +
        L_{matter}(g_{\mu\nu},\varphi^m)
\end{equation}

where $L_{GR}$ denotes the general Lagrangian of general relativity, $L_{matter}$ some general covariant matter Lagrangian compatible with general relativity, and the constant matrix $\gamma_{\alpha\beta}$ is defined by two constants $\Xi,\Upsilon$ as $\mbox{diag}\left\{\Upsilon,-\Xi,-\Xi,-\Xi\right\}$.

\end{theorem}

\begin{proof} The metric $g_{\mu\nu}(x)$ is defined in the preferred coordinates $\x^\a$ by equation \eqref{gdef}. It has signature $(1,3)$. This follows from $\rho>0$ (axiom \ref{afirst}) and negative definiteness of $p^{ij}$ (axiom \ref{apressure}). Because the $t^{\mu\nu}$ transform in the same way as the $g^{\mu\nu}\sqrt{-g}$, the dependence on the preferred coordinates $\x^\a$ of the Lagrangian provided by axiom \ref{aLagrange} remains to be explicit if we replace the $t^{\mu\nu}$ by $g^{\mu\nu}\sqrt{-g}$. Now, the key observation is what happens with our four classical conservation laws during this change of field variables.  They (essentially by construction) become the harmonic condition for the metric $g_{\a\b}(\x)$:
\begin{equation}
\partial_\gamma (g^{\alpha\gamma}\sqrt{-g}) = 
\partial_\gamma (g^{\beta\gamma}\sqrt{-g}) \partial_\beta \x^\alpha(\x)
\equiv \square \x^\alpha(\x) = 0 
\end{equation}
where $\square$ denotes the harmonic operator of the metric $g_{\a\b}(\x)$.  We define now the constants $\Upsilon,\Xi$ as the proportionality factors between the energy and momentum conservation laws and the continuity resp. Euler equations (axioms \ref{acontinuity}, \ref{aEuler}).  For convenience a common factor $(4\pi G)^{-1}$ is introduced. The operator $\square$ is covariant, so that $\square \x^\a(x)$ holds in general coordinates too. So we obtain
\begin{eqnarray}
\frac{\delta S}{\delta\t}   &=& - (4\pi G)^{-1} \Upsilon \square\t(x)\\
\frac{\delta S}{\delta \x^i} &=&   (4\pi G)^{-1} \Xi \square \x^i(x)
\end{eqnarray}
Defining the diagonal matrix $\gamma_{\alpha\beta}$ by $\gamma_{00}=\Upsilon,\gamma_{ii}=-\Xi$, we can write these equations in closed form as \footnote{Note that the coefficients $\gamma_{\alpha\beta}$ are only constants of the Lagrange density, the indices enumerate the scalar fields $\x^\alpha$.  They do not define any fundamental, predefined object of the theory.  Instead, variables of the Lagrange formalism, by construction, are only $g_{\mu\nu}$, $\varphi^m$ and $\x^\alpha$.}
\begin{equation}\label{deltaS}
\frac{\delta S}{\delta \x^\alpha} \equiv-(4\pi G)^{-1}\gamma_{\alpha\beta}\square \x^\beta(x)
\end{equation}

Let's find now the general form of a Lagrangian which gives these equations. The general solution of \eqref{deltaS} is the sum of a particular solution of \eqref{deltaS} and the general solution of the corresponding homogeneous problem. It is easy to find a particular solution $L_0$:
\begin{equation}\label{defL0}
L_{0} = -(8\pi G)^{-1}\gamma_{\alpha\beta}\x^\alpha_{,\mu}\x^\beta_{,\nu} g^{\mu\nu}\sqrt{-g}
\end{equation}
Then we have to consider the difference $L-L_{0}$. Together with $L$ and $L_0$ it's dependence on the preferred coordinates is explicit. Therefore we obtain the corresponding homogeneous problem:
\begin{equation}
\frac{\delta \int(L-L_{0})}{\delta \x^\alpha} \equiv 0 
\end{equation}
But this is simply the condition of general covariance of theorem \ref{Noether2}. Thus, the general solution of the homogenous problem gives most general Lagrangian of general relativity.  So we obtain:
\begin{equation}
L = -(8\pi G)^{-1}\gamma_{\alpha\beta}\x^\alpha_{,\mu}\x^\beta_{,\nu} g^{\mu\nu}\sqrt{-g} + L_{GR}(g_{\mu\nu}) + L_{matter}(g_{\mu\nu},\varphi^m).
\end{equation}
\end{proof}

It should be noted that the ``most general'' Lagrangian means that higher order derivatives of the metric and non-minimal interactions with matter fields are allowed. We could have added some additional requirements restricting, for example, the number of derivatives, to obtain the standard Einstein-Hilbert Lagrangian of general relativity, but see no reason to do this. Instead, we follow the view expressed by Weinberg \cite{Weinberg}: ``I don't see any reason why anyone today would take Einstein's general theory of relativity seriously as the foundation of a quantum theory of gravitation, if by Einstein's theory is meant the theory with a Lagrangian density given by just the term $\sqrt{g}R/16\pi G$.  It seems to me there's no reason in the world to suppose that the Lagrangian does not contain all the higher terms with more factors of the curvature and/or more derivatives, all of which are suppressed by inverse powers of the Planck mass, and of course don't show up at energy far below the Planck mass, much less in astronomy or particle physics.  Why would anyone suppose that these higher terms are absent?''

Another point worth to note is that the signs and values of the constants $\Upsilon, \Xi$ remain undefined. One could use a linear transformation $\t(x)\to c_{\t} \t(x)$, $\x^i(x) \to c_{\x} \x^i(x)$ to normalize them to $\abs{\Upsilon} = \abs{\Xi} = 1$, which corresponds to some choice of natural units for the background coordinates. But, given that the values of these natural units are unknown, for comparison with observations it seems easier to use harmonic coordinates connected with our units, and to leave the constants $\Upsilon, \Xi$ arbitrary, to be defined by observation.

The choice of the signs of $\Upsilon$ and $\Xi$ is also not restricted by the derivation. One may think that the sign has to be defined as appropriate for the similar term with scalar fields $U^\a(x)$ instead of the $\x^\a$. But this is wrong. The coordinates $\x^\a(x)$ follow very different initial and boundary conditions, and the sign which would be appropriate for a scalar fields $U^\a(x)$ may be the wrong choice (cf. app. \ref{appGRCF}). Thus, the choice has to be left to more detailed considerations of stability. In fact, considering the gravitational collapse as well as the big bang, we find that using the ``wrong'' sign for $\Upsilon$, namely $\Upsilon>0$, prevents the standard GR big bang and black hole singularities (see app. \ref{appPredictions}).

\section{Lorentz Ether Theory as a Metric Theory of Gravity}

Now, the explicit formalism has been quite useful to formulate the axioms of GLET and to derive the effective Lagrangian in theorem \ref{main}.  But it is only a technical formalism.  Once theorem \ref{main} has been proven, we can as well return to the usual, implicit non-covariant formulation of the theory.

\begin{theorem}
Assume we have a theory which fulfills axioms \ref{afirst}-\ref{alast}.  Then formula \eqref{gdef} defines an isomorphism between solutions of this theory and the solutions of the (non-covariant) metric theory of gravity defined by the Lagrangian
\begin{equation}
L_{GLET} = L_{GR} + L_{matter}
  - (8\pi G)^{-1}(\Upsilon g^{00}-\Xi \delta_{ij}g^{ij})\sqrt{-g}
\end{equation}
and the additional ``causality condition'' $g^{00}>0$.
\end{theorem}
\begin{proof}
Indeed, the Lagrangian $L_{GLET}$ is simply the Lagrangian $L$ of theorem \ref{main} written in the preferred coordinates $\x^\alpha$. It follows from \eqref{gdef} and $\rho>0$ that for solutions of GLET the causality condition $g^{00}>0$ is fulfilled.  For a given solution of GLET formula \eqref{gdef} also allows to define the pre-image solution in terms of $\rho, v^i, p^{ij}$. 
\end{proof}

Thus, our class of condensed matter theories is equivalent to a metric theory of gravity with a Newtonian framework as a predefined background.  As equations of the theory in the preferred coordinates we obtain the Einstein equations of general relativity with two additional terms:
\begin{equation}\label{GLETeq}
G^\a_\b = 8\pi G (T_m)^\a_\b
   + (\Lambda +\gamma_{\gamma\delta}g^{\gamma\delta}) \delta^\a_\b
   - 2g^{\a\gamma}\gamma_{\gamma\b}
\end{equation}
as well as the harmonic equations -- the conservation laws
\begin{equation}
\partial_\b  (g^{\b\alpha}\sqrt{-g}) =  0. 
\end{equation}

Note that the ``energy-momentum tensor of matter'' $(T_m)^\a_\b$ should not be mingled with the classical energy-momentum tensor $t^\a_\b$ of the ether defined in axiom \ref{afirst}, which is the full energy-momentum tensor.  There is also another form of the conservation laws. The basic equation may be simply considered as a decomposition of the full energy-momentum tensor $g^{\a\gamma}\sqrt{-g}$ into a part $(T_m)^\a_\b$ which depends on matter fields, and a part $(T_g)^\mu_\nu$ which depends on the gravitational field. Thus, for
\begin{equation}
(T_g)^\a_\b = (8\pi G)^{-1}\left(\delta^\a_\b(\Lambda
              + \gamma_{\gamma\delta}g^{\gamma\delta})
              - G^\a_\b\right)\sqrt{-g}
\end{equation}
we obtain a second form of the conservation law
\begin{equation}
\partial_\mu  ((T_g)^\a_\b +  (T_m)^\a_\b\sqrt{-g}) = 0.
\end{equation}
Thus, we have even two equivalent forms for the conservation laws, one with a subdivision into gravitational and matter part, and another one where the full tensor depends only on the gravitational field. In the covariant formalism, the equations are
\begin{equation}\label{GLETcov}
\begin{split}
G^\mu_\nu(x)  &= 8\pi G (T_m)^\mu_\nu(x)
   + (\Lambda +\gamma_{\alpha\beta}\x^\alpha_{,\kappa}(x)\x^\beta_{,\lambda}(x)g^{\kappa\lambda}(x)) \delta^\mu_\nu\\
   &- 2g^{\mu\kappa}(x)\gamma_{\alpha\beta}\x^\alpha_{,\kappa}(x)\x^\beta_{,\nu}(x)
\end{split}
\end{equation}
and
\begin{equation}
\square \x^\alpha(x) = \partial_\mu  (g^{\mu\beta}(x)\sqrt{-g}) \partial_\beta \x^\alpha(x) =  0. 
\end{equation}

\section{Internal Observers and the Einstein Equivalence Principle}

The explicit form of the Lagrangian is, if we replace the preferred coordinates $\x^\mu$ with scalar fields $U^\mu$, formally equivalent to the Lagrangian of ``general relativity with clock fields'' (GRCF) considered by Kuchar and Torre \cite{Kuchar} in the context of GR quantization in harmonic gauge:

\begin{equation}
L_{GRCF} = -(8\pi G)^{-1}\gamma_{\alpha\beta}U^\alpha_{,\mu}U^\beta_{,\nu}
        g^{\mu\nu}\sqrt{-g} + L_{GR}(g_{\mu\nu}) +
        L_{matter}(g_{\mu\nu},\varphi^m).
\end{equation}

The only difference is that in GRCF we have scalar fields $U^\alpha(x)$ instead of preferred coordinates $\x^\alpha(x)$.  But this makes above theories completely different as physical theories. We show this in detail in app. \ref{appGRCF}.  Despite these differences, internal observers of our condensed matter theories have a very hard job to distinguish above theories by observation:

\begin{corollary} 
In GLET internal observers cannot falsify GRCF by any local observation.\footnote{Note that in the reverse direction the corollary doesn't hold: There are local solutions of GRCF where the fields $U^\mu(x)$ cannot be used as coordinates even locally.}
\end{corollary}

\begin{proof}Indeed, internal observers are creatures described by field configurations of the fields $\left(\rho(\x,\t), v^i(\x,\t), p^{ij}(\x,\t), \varphi^m(\x,\t)\right)$.  Whatever they observe may be described by a solution $\left(\rho(\x,\t), v^i(\x,\t), p^{ij}(\x,\t), \varphi^m(\x,\t)\right)$.  Now, following theorem \ref{main}, they can describe all their experiments and observations also as local solutions of GRCF for some ``fields'' $g^{\mu\nu}, \varphi^m(x), U^i(x), U^0(x)$.  \end{proof}

Based on this observation we can prove now the Einstein Equivalence Principle.  For this purpose we have to clarify what is considered to be a matter field.  This question is theory-dependent, and there are two possibilities for this: First, the internal observers of GLET may believe into GRCF.  In this case, they consider the four special scalar fields $\x^\alpha(x)$ as non-special scalar matter fields together with the other matter fields $\varphi^m(x)$.  Second, they may believe into GLET.  Then they consider only the fields $\varphi^m(x)$ as non-gravitational matter fields.  This gives two different notions of non-gravitational experiments and therefore different interpretations of the EEP. But in above interpretations of the EEP it holds exactly:

\begin{corollary} For internal observers of a condensed matter theory GLET which fulfills axioms \ref{afirst}-\ref{alast} the Einstein Equivalence Principle holds in above interpretations.
\end{corollary}

\begin{proof} Indeed, GRCF is a variant of general relativity where the EEP holds exactly.  Because the EEP is a local principle (it holds if it holds for all local observations) and internal observers cannot falsify GRCF by local observations, they cannot falsify the EEP by observation. Thus, in the first interpretation the EEP holds.  In the second case, only the $\varphi^m$ are interpreted as matter fields. Nonetheless, we have also a metric theory of gravity coupled with the matter fields $\varphi^m$ in the usual covariant way, and therefore the EEP holds.\end{proof}

\section{The GR Limit}

In the limit $\Xi,\Upsilon\to 0$ we obtain the Lagrangian of general relativity.  Thus, we obtain the classical Einstein equations. Nonetheless, some points are worth to be discussed here:

\subsection{The Remaining Hidden Background}

In this limit the absolute background becomes a hidden variable.

Nonetheless, the theory remains different from general relativity even in this limit.  The hidden background leads to some important global restrictions: Non-trivial topologies are forbidden, closed causal loops too.  There always exists a global harmonic time-like function. \footnote{There is a general prejudice against such hidden variables. It is sometimes argued that a theory without such restrictions should be preferred as the ``more general'' theory.  But this prejudice is not based on scientific methodology as described by Popper's criterion of empirical content.  If this limit of GLET is true, an internal observer cannot falsify GR in the domain of applicability of this limit.  On the other hand, if GR is true, GLET is falsifiable by observation: If we observe non-trivial topology, closed causal loops, or situations where no global harmonic time exists, we have falsified GLET.  In this sense, the GR limit of GLET should be preferred by Popper's criterion of predictive power in comparison with GR.}

\subsection{The Absence of Fine Tuning in the Limit}

Note that we need no fine-tuning to obtain the GR limit $\Xi,\Upsilon\to 0$.  All non-covariant terms depend only on the metric $g^{\mu\mu}\sqrt{-g}$, not on its derivatives $\partial_\kappa g_{\mu\nu}$.  That means, in a region where we have well-defined upper and lower bounds $0<c^\mu_{min}<|g^{\mu\mu}\sqrt{-g}|<c^\mu_{max}$ (in the preferred coordinates) all we have to do is to consider low distance high frequency effects.  In comparison with global cosmology, solar system size effects obviously fit into this scheme.  Thus, the non-covariant terms may be reasonably considered as cosmological terms, similar to Einstein's cosmological constant $\Lambda$.

On the other hand, there may be effects in solar-system-like dimension where these cosmological terms cannot be ignored: In the preferred coordinates, the terms $g^{\mu\mu}\sqrt{-g}$ may become large.  An example is the region near horizon formation for a collapsing star. As shown in app. \ref{appPredictions}, in the case of $\Upsilon>0$ this term prevents black hole formation, and we obtain stable ``frozen star'' solutions.

\subsection{Comparison with Other Approaches to GR}

Part of the beauty of GR is that there many different ways to general relativity.  First, there are remarkable formulations in other variables (ADM formalism \cite{ADM}, tetrad, triad, and Ashtekar variables \cite{Ashtekar}) where the Lorentz metric $g_{\mu\nu}$ appears as derived.  Some may be considered as different interpretations of general relativity (like ``geometrodynamics'').  On the other hand we have theories where the Lorentz metric is only an effective metric and the Einstein equations appear in a limit (spin two field in QFT on a standard Minkowski background \cite{Feynman} \cite{Weinberg65} \cite{Deser}, string theory \cite{Polchinski}), Sakharov's approach \cite{Sakharov}.

In this context, the derivation of GLET, combined with the limit $\Xi,\Upsilon\to 0$, may be considered as ``yet another way to GR''. In this interpretation, it should be classified as part of the second group of derivations: The metric is, as in these approaches, not fundamental.  Moreover, a flat hidden background remains.

\section{The No-Gravity Limit}

Last but not least, let's consider shortly the no gravity limit of GLET. For a metric theory of gravity, it is the Minkowski limit $g_{\a\b}(\x)=\eta_{\a\b}$.  In this limit, the additional terms of GLET in comparison with GR no longer vary, therefore, may be omitted. Thus, the no gravity limit of GLET coinsides with the no gravity limit of GR, and is, therefore, equivalent to SR.

In this limit, we have 

\begin{eqnarray*}
\rho(\x)  &=  &\rho_0 \\
v^i(\x)   &=  &0 \\
p^{ij}(\x)&=  &p_0 \delta^{ij}
\end{eqnarray*}

with constants $\rho_0>0$, $p_0<0$.  Thus, the ether is static and incompressible.  It follows from the properties of the no gravity limit of the matter Lagrangian, which is identical in GR and GLET, that clocks and rulers behave as they behave in SR.  

Thus, in the no gravity limit we obtain a static and incompressible ether in a Newtonian framework. The remarkable property of the Lagrangian of the remaining fields, which describe the remaining material properties of the ether, is, that, nonetheless, the Einstein equivalence principle for them holds. As a consequence, the Lagrangian has to follow the general rules for Lagrangians in special relativity, in particular, has to be Lorentz-covariant. 

The consequence of the Lorentz-covariance of the matter Lagrangian is that clocks and rulers constructed from matter show time dilation and length contraction in agreement with the formulas of special relativity. Especially moving clocks have to be slower, and moving rulers have to be contracted, in such a way that the preferred time coordinate $T$ remains unobservable. These properties of the no gravity limit of GLET are the properties of the classical Lorentz ether. This justifies the interpretation of GLET as a generalization of Lorentz ether theory to gravity.

Nonetheless, it seems worth to note a minor but important difference: In the classical ether theories, the ether has been considered only as a substance which explains electromagnetic waves.  Usual matter has been considered to be something different.  In GLET, a key point of the derivation is that all fields -- including the fields which describe fermionic particles -- describe degrees of freedom of the ether itself.  Thus, there is no type of matter which ``interacts with the ether''. There is only a single universal ether, and nothing else. To talk about ``matter interacting with the ether'' is as meaningless as to talk about water waves interacting with water.

\section{Conclusions}

We have presented a new metric theory of gravity starting with a set of axioms for a classical condensed matter theory in a Newtonian background.  Relativistic symmetry (the Einstein equivalence principle) is derived here: Internal observers of the medium (called ``general Lorentz ether'') are unable to distinguish all properties of their basic medium.  All their local observations fulfill the Einstein equivalence principle.

While we have used an unorthodox formalism -- explicit dependence of a non-covariant Lagrangian on the preferred coordinates -- the basic principles we have used in the axioms are well-known and accepted: continuity and Euler equations, Lagrange formalism, and Noether's theorem, which relates them with translational symmetries.

The differences between GR and our theory are interesting: The new theory predicts inflation and a stop of the gravitational collapse, preventing black hole and big bang singularities.  As a condensed matter theory in a Newtonian framework, classical canonical quantization concepts known from quantum condensed matter theory may be applied to quantize the theory.

The new theory is able to compete with general relativity even in the domain of beauty: It combines several concepts which are beautiful already by them-self: Noether's theorem, conservation laws, harmonic coordinates, ADM decomposition.  Instead of the ugly situation with local energy and momentum of the gravitational field in GR, we have nice, local conservation laws.  Moreover we have no big bang and black hole singularities.

To establish this theory as a serious alternative to general relativity, this paper may be only a first step.  We have to learn more about the predictions of GLET which differ from GR, experimental bounds for $\Xi, \Upsilon$ have to be obtained, the quantization program has to be worked out in detail.  The methodological considerations related with such fundamental notions like EPR realism have to be worked out too.  Nonetheless, the results we have obtained are already interesting enough, and the new theory seems able to compete with general relativity in all domains we have considered -- cosmology, quantization, explanatory power, and even simplicity and beauty.

\begin{appendix}

\section{About some common misunderstandings}

Many common misunderstandings of the theory are related with the nature of the equations which define the gravitational field:

\begin{eqnarray*}
 \hat{g}^{00} = g^{00} \sqrt{-g} &=  &\rho \\
 \hat{g}^{i0} = g^{i0} \sqrt{-g} &=  &\rho v^i \\
 \hat{g}^{ij} = g^{ij} \sqrt{-g} &=  &\rho v^i v^j + p^{ij}
\end{eqnarray*}

The equation itself is often misrepresented as if it is a physical equation, which describes the interaction of the usual spacetime metric $g_{\mu\nu}(x)$ with the ether, which is something like usual condensed matter. This is wrong. There is no separate gravitational field $g_{\mu\nu}(x)$ which interacts with the ether. Instead, there are two possible sets of field variables we can use to describe one and the same thing -- the ether. One set of variables consists of $\rho(\x),v^i(\x),p^{ij}(\x)$ and other material properties $\varphi_m(\x)$, a set which we know from condensed matter physics as an appropriate choice for the description of condensed matter. The other set of fields consists of $g_{\a\b}(\x)$ and the $\varphi_m(\x)$. The equation describes how these two sets of variables are connected with each other in the preferred coordinates $\x^\a$: It is an algebraic transformation of variables.

The ether density $\rho_{ether}$ has to be clearly distinguished from densities of usual matter $\rho_{matter}$, which appear on the right-hand side of the Einstein equations in the energy-momentum tensor of matter. They are as different as, for comparison, the density of water from the density of waves on its surface, or the density of air from the density of sound waves in it. 

In relation with this definition of the ether density, the question of units may arise. Naively, one may think about the ether density as having the unit $g/cm^3$. The expression on the other side of the equation has the unit $cm^3/s$. Thus, a factor containing appropriate units seems to be missed.

But the density of the ether $\rho$ is a qualitatively new entity. We have nothing to compare it with: We cannot take one piece of ether and measure its weight in $g$. Therefore, to assume that the ether density is measured in $g/cm^3$ is physically nonsensical. We also don't know (yet) the size of the ether atoms, so we cannot count their number in a given volume to measure the ether density in $atoms/cm^3$. What is, in this case, the appropriate unit to measure the ether density in our theory? Physical units should be defined in connection with a way to measure them. But the only way to measure $\rho$ of the ether in our theory is to measure the metric $g_{\a\b}(\x)$ in the preferred coordinates and then to compute $\rho(\x)$ according to the definition $\rho = g^{00} \sqrt{-g}$. Thus, it is the definition, without any additional unit factor, which provides us with a way (the only way) to measure the ether density, and, therefore, allows a definition of a natural unit for the ether density.

The situation would be different if, say, we would know the critical length of some atomic structure of the ether. In this case, there would be a natural unit for the ether density, in $atoms/cm^3$ of the ether. Correspondingly we would need, in this case, an additional coefficient for the transformation between these two natural units for the ether density, which, in analogy to the Avogadro number, could be named ``ether Avogadro number''. But, even in this case, the unit for the ether density given by the definition $\rho = g^{00} \sqrt{-g}$ could survive as well as the unit $mol$ has survived in chemistry. Moreover, even in this case there would be no need to introduce a constant into the definition itself.

Once the unit of $\rho$ is defined appropriately, corresponding units for the ether momentum vector $\rho v^i$ and the ether pressure tensor $p^{ij}$ follow automatically.

\section{Can every Lagrangian be rewritten in explicit form?}
\label{appGeneral}

In axiom \ref{aLagrange} we assume that the Lagrangian is given in explicit form.  Now, the question we want to discuss here is if this is a non-trivial restriction or not. If it is not, then our thesis \ref{thesis:explicit} holds.

In some sense, this can be considered as a corollary to Kretschmann's \cite{Kretschmann} thesis that every physical theory may be reformulated in a general covariant way.  We would have to reformulate a given theory with Lagrangian $F$ into a general covariant form, and when, as far as possible, try to replace all objects used to describe the absolute Newtonian framework in this covariant description by a description which depends explicitly on the preferred coordinates $\x^\alpha(x)$.
\footnote{Remember that there has been some confusion about the role of covariance in general relativity.  Initially it was thought by Einstein that general covariance is a special property of general relativity.  Later, it has been observed that other physical theories allow a covariant description too.  The classical way to do this for special relativity (see \cite{Fock}) is to introduce the background metric $\eta_{\mu\nu}(x)$ as an independent field and to describe it by the covariant equation $R^\mu_{\nu\kappa\lambda}[\eta]=0$.  For Newton's theory of gravity a covariant description can be found in \cite{MTW}, \S 12.4.  The general thesis that every physical theory may be reformulated in a general covariant way was proposed by Kretschmann \cite{Kretschmann}.}

In this sense, let's try to describe how the classical geometric objects related with the Newtonian framework may be described in an explicit way:

\begin{itemize}

\item The preferred foliation: It is immediately defined by the coordinate $\t(x)$.

\item The Euclidean background space metric: A scalar product $(u,v)=\delta_{ij}u^iv^j$ may be presented in explicit form as $(u,v)=\delta_{ij}\x^i_{,\mu}\x^j_{,\nu}u^\mu v^\nu$.

\item Absolute time metric: Similarly, the (degenerate) metric of absolute distance in time may be defined by $(u,v)=\t_{,\mu}\t_{,\nu}u^\mu v^\nu$.

\item The preferred coordinates also define tetrad and cotetrad fields: $d\x^\alpha = \x^\alpha_{,\mu} dx^\mu$, for the vector fields $\partial/\partial \x^\alpha = (J^{-1})^\mu_\alpha(\x^\beta_{,\nu})\partial/\partial x^\mu$ we need the inverse Jacobi matrix $J^{-1}$, which is a rational function of the $\x^\beta_{,\nu}$.

\item Arbitrary upper tensor components $t^{\alpha_1,..\alpha_n}$ may be transformed into $\x^{\alpha_1}_{,\mu_1}\cdots \x^{\alpha_n}_{,{\mu_n}}t^{\mu_1,..\mu_n}$.  For lower indices we have to use again the inverse Jacobi matrix $J^{-1}$.

\item Together with the background metrics we can define also the related covariant derivatives.

\end{itemize}

Now, these examples seem sufficient to justify our hypothesis.

But it should be mentioned that there is no algorithm which allows to transform a given non-covariant theory into a form with explicit dependence on the preferred coordinates. We have to distinguish here two meanings of a ``non-covariant theory''. One meaning is a theory where the rules of transformation into other systems of coordinates are well-defined, but it appears that the laws of the non-covariant theory change if we change the coordinates. This is the type of non-covariance where we can hope to obtain an algorithm for transforming it into a variant with explicit dependence: The theory is already described in terms of objects with well-defined transformation rules.

The situation is different for theories which are defined only in a single preferred system of coordinates, without any rules how to rewrite the equations of the theory in other systems of coordinates. For example, for theories containing some function like $\rho(\x,\t)$ in the preferred coordinates it is not clear at all how to rewrite it in other coordinates. There are different possibilities: $\rho(\x,\t)$ may transform like a scalar field, it may transform, taken together with the $\rho(\x,\t)v^i(\x,\t)$, like a component of a four-vector field, and it may transform, following \eqref{gdef}, like a component of a symmetric tensor field. Other, more complex possibilities cannot be excluded as well. Which of these choices is the correct one? In addition, if we use our choice \eqref{gdef} for transforming $\rho(\x,\t)$, using the pressure tensor $p^{ij}(\x,\t)$, the fact that the Euler equations define $p^{ij}(\x,\t)$ only modulo a constant $p^{ij}(\x,\t)\to p^{ij}(\x,\t)+C\delta^{ij}$ leads to another non-uniqueness: Which is the correct choice of this constant? These questions cannot be answered uniquely if the equations of the theory are defined only in the preferred coordinates.

Let's note again that our hypothesis is not essential for our derivation.  Simply, if the hypothesis is false, our assumption that the Lagrangian is given in explicit form (which is part of axiom (\ref{aLagrange})) is a really non-trivial restriction. This would be some loss of generality and therefore of beauty of the derivation, but in no way fatal for the main results of the paper.  The considerations given here already show that the class of theories which allows an explicit formulation is quite large.

\section{Justification of the Euler-Lagrange Equations for the Preferred Coordinates} \label{appCoordinates}

In the Euler-Lagrange formalism for a Lagrangian with explicit dependence on the preferred coordinates we propose to handle the preferred coordinates like usual fields.  Of course, every valid set of coordinates $\x^\alpha(x)$ defines a valid field configuration.  But the reverse is not true.  To define a valid set of coordinates, the functions $\x^\alpha(x)$ have to fulfill special local and global restrictions: the Jacobi matrix should be non-degenerated everywhere, and the functions should fulfill special boundary conditions.  In app. \ref{appGRCF} we find that this makes the two theories different as physical theories.

Here we want to prove theorem \ref{Euler-Lagrange} that nonetheless the Lagrange formalism works as usual.  The description of the Lagrange function in explicit form is, in comparison with an implicit description, only another way to describe the same minimum problem. To solve this minimum problem we can try to apply the standard variational calculus.  Especially we can also vary the preferred coordinates $\x^\alpha$, which are, together with the other fields, functions of the manifold.  But there is a subtle point which has to be addressed.  The point is that not all variations $\delta \x^\alpha(x)$ are allowed -- only the subset with the property that $\x^\alpha(x)+\delta \x^\alpha(x)$ defines valid global coordinates. Therefore, to justify the application of the standard Euler-Lagrange formalism we have to check if this subset of allowed variations is large enough to give the classical Euler-Lagrange equations.

Fortunately this is the case.  To obtain the Euler-Lagrange equations we need only variations with compact support, so we don't have a problem with the different boundary conditions.  Moreover, for sufficiently smooth variations ($\delta \x^\alpha \in C^1({\mathbb R}^4)$ is sufficient) there is an $\varepsilon$ so that $\x^\alpha(x)+\varepsilon\delta \x^\alpha(x)$ remains to be a system of coordinates: indeed, once we have an upper bound for the derivatives of $\delta \x^\alpha$ (because of compact support), we can make the derivatives of $\varepsilon\delta \x^\alpha$ arbitrary small, especially small enough to leave the Jacobi matrix of $\x^\alpha+\varepsilon\delta \x^\alpha$ non-degenerated.  Such sufficiently smooth variations are sufficient to obtain the Euler-Lagrange equations. 

Now, to obtain the Euler-Lagrange equations we need only small variations.  Thus, the geometrical restrictions on global coordinates do not influence the derivation of the Euler-Lagrange equations for the $\x^\alpha(x)$.  For the preferred coordinates we obtain usual Euler-Lagrange equation, as if they were usual fields, despite their special geometric nature.

\section{Discussion of the Derivation}
\label{appDerivation}

The extremely simple derivation of exact general-relativistic symmetry in the context of a classical condensed matter theory looks very surprising.  It may be suspected that too much is hidden behind the innocently looking relation between Lagrange formalism and the condensed matter equations, or somewhere in our unorthodox ``explicit formalism''.  Therefore, let's try to understand on a more informal, physical level what has happened, and where we have made the physically non-trivial assumptions which lead to the very non-trivial result -- the EEP.

A non-trivial physical assumption is, obviously, the classical Euler equation.  This equation contains non-trivial information -- that there is only one medium, which has no interaction, especially no momentum exchange, with other media.  This is, indeed, a natural but very strong physical assumption.  And from this point of view the derivation looks quite natural: What we assume is a single, universal medium.  What we obtain with the EEP is a single, universal type of clocks.  An universality assumption leads to an universality result.

From point of view of parameter counting, all seems to be nice too. We have explicitly fixed four equations, and obtain an independence from four coordinates.  All the effective matter fields are by construction fields of a very special type -- inner degrees of freedom, material properties, of the original medium.  Therefore it is also not strange to see them closely tied to the basic degrees of freedom (density, velocity, pressure) which define the effective gravitational field.

The non-trivial character of the existence of a Lagrange formalism seems also worth to be mentioned here.  A Lagrange formalism leads to a symmetry property of the equation -- they should be self-adjoint. This is the well-known symmetry of the ``action equals reaction'' principle.  It can be seen where this symmetry has been applied if we consider the second functional derivatives.  The EEP means that the equations for effective matter fields $\varphi^m$ do not depend on the preferred coordinates $\x^\alpha$.  This may be formally proven in this way:

\begin{equation} 
   \frac{\delta}{\delta \x^\a}     \frac{\delta S}{\delta \varphi^m}
 = \frac{\delta}{\delta \varphi^m} \frac{\delta S}{\delta \x^\a}
 = \frac{\delta}{\delta \varphi^m} \mbox{[cons. laws]} = 0 
\end{equation}

Thus, we have applied here the ``action equals reaction'' symmetry of the Lagrange formalism and the property that the fundamental classical conservation laws (connected with translational symmetry by Noether's theorem) do not depend on the material properties $\varphi^m$.

Our considerations seem to indicate that the relativistic symmetry (EEP) we obtain is explained in a reasonable way by the physical assumptions which have been made, especially the universality of the medium and the special character of the ``effective matter'' degrees of freedom -- assumptions which are implicit parts of the Euler equation -- and the ``action equals reaction'' symmetry of the Lagrange formalism.

It seems also necessary to clarify that the states which cannot be distinguished by local observation of matter fields are not only different states, they can be distinguished by gravitational experiments.  The EEP holds (we have a metric theory of gravity) but not the Strong Equivalence Principle (SEP).  The SEP holds only in the GR limit.  For the definitions of EEP and SEP used here see, for example, \cite{Will}.

\section{What Happens During Evolution with the Causality Condition?}
\label{appFailure}

We have proven that solutions of $GRCF$ which are images of solutions of GLET fulfill some restrictions (time-like $\t(x)$, $\x^\alpha(x)$ are coordinates). Now, it is reasonable to ask if these restrictions are compatible with the evolution equations of GRCF.  Assume that we have a solution of GRCF which fulfills all necessary restrictions for some time $\t(x)< \t_0$. This solution now evolves following the evolution equations of GRCF. Is there any warranty that these global restrictions remain to be fulfilled also for $\t(x)\ge \t_0$?  The answer is no. There is not such general warranty. But, as we will see, there is also no necessity for such a warranty.

For a given Lorentz ether theory (with fixed material Lagrangian) such a warranty would follow from a global existence and uniqueness theorem.  Indeed, having such a theorem for the condensed matter theory, it would be sufficient to consider the image of the global solution. \footnote{In this context it seems not uninteresting to note that the famous local existence and uniqueness theorems for GR given by Choquet-Bruhat \cite{Choquet-Bruhat} are based on the use of harmonic gauge.  That means, they may be interpreted as local existence and uniqueness theorems for the limit $\Xi,\Upsilon\to 0$ of $GLET$, combined with considerations that local existence and uniqueness for this limit (in this gauge) is the same as local existence and uniqueness for general relativity.}

But in such a general class of theories as defined by our axioms \ref{afirst}-\ref{alast} where is no hope for general global existence theorems and, as a consequence, for a warranty that the restrictions remain valid for $\t(x)>\t_0$.  Moreover, theories without global existence theorems are quite reasonable as condensed matter theories. It happens in reality that some material tears.  In this case, the continuous condensed matter theory which is appropriate for this material should have a limited domain of application, and there should be solutions which reach $\rho=0$ somewhere.  But according to equation \eqref{gdef} $\rho=0$ corresponds to a violation of the condition that $\t(x)$ is time-like.  Thus, if such a material is described by a theory which fulfills our axioms, and a solution describes a state which at some time starts to tear, the restriction that $\t(x)$ is time-like no longer holds.

\section{Reconsideration of Old Arguments Against The Ether}
\label{appOldArguments}

The generalization of Lorentz ether theory we have presented here removes some old, classical arguments against the Lorentz ether -- arguments which have justified the rejection of the Lorentz ether in favor of relativity:

\begin{itemize}

\item There was no viable ether theory of gravity. Now, we have found an ether theory of gravity with GR limit which seems viable;

\item The assumptions about the Lorentz ether have been ad hoc. There was no explanation of relativistic symmetry, the relativistic terms have had ad hoc character. Instead, the assumptions we make for our medium seem quite natural. In particular, and we {\em derive\/} the EEP instead of postulating it. As a consequence, the relativistic lenght contraction and time dilation which has to be postulated in the Lorentz ether can now be derived from other principles.

\item There was no explanation of the general character of relativistic symmetry. The theory was only electro-magnetic. Instead, in the new concept the ``ether'' is universal, all matter fields describe properties of the ether. This, in particular, explains the universality of the EEP and the gravitational field;

\item There was a violation of the ``action equals reaction'' principle: The matter was influenced by the ether, but there was no reverse influence of matter on the stationary and incompressible ether. Now we have a Lagrange formalism. This guarantees the ``action equals reaction'' principle and gives a compressible, instationary ether;

\item There were no observable differences in the predictions of the Lorentz ether and special relativity --- there are important and interesting differences between our theory and general relativity (see appendix \ref{appPredictions}).

\end{itemize}

\subsection{What remains: The unobservability of the preferred frame}\label{appUnobservable}

But one classical argument against the Lorentz ether remains valid: The preferred frame remains unobservable. Indeed, the GR part of the Lagrangian has general covariance. The only term which does not have general covariance --- the \eqref{defL0} term --- depends on the $\x^\a$ only via the expression $\gamma_{\mu\nu}(x) = \gamma_{\a\b}\x^\a_{,\mu}(x)\x^\b_{,\nu}(x)$.  Thus, transformations of the coordinates which leave $\gamma_{\mu\nu}(x)$ invariant leave the whole Lagrangian invariant.

And, whatever the choices of the constants, among the transformations which leave the Lagrangian invariant will be also transformations which mingle absolute time $\t(x)$ with the spatial coordinates $\x^i(x)$. As a consequence, it is not possible to measure the true time $\t(x)$.

This situation is quite close to the situation of the Lorentz ether in special relativity. But here the theory which plays the role of special relativity is not general relativity, but a theory of gravity with preferred Minkowski background ---  Logunov's ``relativistic theory of gravity'' (RTG) \cite{Logunov1}.  The relation between these two theories will be considered in more detail in appendix \ref{RTG}.

It should be noted that what we name here ``metaphysical'' is metaphysical only at the level of a continuous field theory. Given the condensed matter interpretation, the natural development of GLET is a discrete (atomic) ether model. Once GLET covers only a few general fields --- density, velocity and pressure --- the discrete model should be based on the theory of the other, material properties of the ether.  In case of our universe, these fields should give the standard model of particle physics in some limit. Such a model has been proposed in \cite{clm}. This model definitely depends on the choice of a preferred frame, and, if it would be possible to measure effects of the critical length of this model, we would see violations of relativistic symmetry.

As a consequence, this ether model for the standard model of particle physics is incompatible with RTG. But it is, at least at the metaphysical level, compatible with GLET. In other words, these purely metaphysical differences at the level of continuous field theory become physical differences at the level of an atomic ether model.

The purely metaphysical differences between the two theories also lead to completely different quantization programs (see app. \ref{appQuantization}).

\section{Non-Equivalence of General Lorentz Ether Theory and General Relativity with Clock Fields}
\label{appGRCF}

In theorem \ref{main} we have proven that solutions of GLET also fulfill the equations of general relativity with four additional scalar fields $\x^\alpha(x)$ which do not interact in any way with the other matter.  In the context of GR quantization in harmonic gauge, this Lagrangian has been considered by Kuchar and Torre \cite{Kuchar}.  Following them, we name it ``general relativity with clock fields'' (GRCF).

Now, a very important point is that, despite theorem \ref{main}, GLET is not equivalent to GRCF as a physical theory.  Indeed, what we have established in theorem \ref{main} is only a very special relationship: for solutions of GLET we obtain ``image solutions'' which fulfill the \emph{equations} of GRCF.  But there are a lot of important differences which make above theories different from physical point of view:

\begin{itemize}

\item \emph{Different notions of completeness:} A complete, global solution of GLET is defined for all $-\infty<\x^i,\t<\infty$.  It is in no way required that the effective metric $g_{\a\b}(\x)$ of this solution is complete.  Therefore, the image of this solution in GRCF is not necessarily a complete, global solution of GRCF.

\item \emph{Global restrictions of topology:} GRCF has solutions with non-trivial topo\-logy. Instead, every image of a solution of GLET has trivial topology.

\item \emph{Global hyperbolicity:} GRCF has solutions which are not globally hyperbolic, especially solutions with closed causal loops.  Instead, every image of a solution of GLET is globally hyperbolic.  Moreover, there exists a global harmonic time-like function: $\t(x)$ on images.

\item \emph{Special character of $\x^i,\t$:} GRCF contains a lot of solutions so that the fields $\x^i(x),\t(x)$ do not define a global system of coordinates. Instead, for every image of a solution of GLET the fields $\x^i(x),\t(x)$ have this property.

\item \emph{Physically unreasonable boundary conditions:} Physically reasonable solutions of GRCF are only solutions with reasonable boundary conditions in infinity.  In particular, a reasonable boundary condition is that all physical fields have some upper bounds.  This leads to $|\x^i(x)|,|\t(x)|<C$ for some constant $C$.  Instead, bounds of this type have to be violated by all coordinates $\x^\a(x)$. Thus, no image of a GLET solution defines a global physically reasonable solution of GRCF.

\end{itemize}

Therefore, the two theories are conceptually very different theories --- they don't share even a single global physically reasonable solution. They only look similar if we ignore the very special geometric nature of the preferred coordinates $\x^i,\t$ of GLET.

\section{Comparison with RTG}\label{RTG}

There is also another theory with the same Lagrangian -- the ``relativistic theory of gravity'' (RTG) proposed by Logunov et al. \cite{Logunov,Logunov1,Logunov2}.  In this theory, we have a Minkowski background metric $\eta_{\mu\nu}$.  The Lagrangian of RTG is

\[L = L_{Rosen} + L_{matter}(g_{\mu\nu},\psi^m)
    - \frac{m_g^2}{16\pi}(\frac{1}{2}\eta_{\mu\nu}g^{\mu\nu}\sqrt{-g}
            - \sqrt{-g} - \sqrt{-\eta})
\]

If we identify the Minkowski coordinates in RTG with the preferred coordinates in GLET, the Lagrangians are equivalent as functions of $g_{\mu\nu}$ for the following choice of constants:
        $\Lambda=-\frac{m_g^2}{ 2}<0$,
        $\Xi=-\eta^{11}\frac{m_g^2}{ 2}>0$,
        $\Upsilon=\eta^{00}\frac{m_g^2}{ 2}>0$.
In this case, the equations for $g^{\mu\nu}$ coincide. The harmonic equation for the Minkowski coordinates hold in RTG \cite{Logunov1,Logunov2}.  As a consequence, the equations of the theories coincide. This holds despite the fact that different fields are considered as physical fields, so that, at a first look, the covariant versions seem to be different. In the covariant RTG version, we have equations for the metric $\eta_{\mu\nu}(x)$, while in our theory we have equations for the coordinates $\x^\mu(x)$, while the $\eta_{\mu\nu}$ appear only as coefficients. But the two versions are connected by the standard formula for the metric in different coordinates: $\eta_{\mu\nu}(x) = \eta_{\alpha\beta}\x^\alpha_{,\mu}(x)\x^\beta_{,\nu}(x)$. 

The two theories are very close to each other: In particular, their equations are identical. Above theories are metric theories of gravity with a preferred background. The main difference is that this preferred background has, in case of RTG, a Minkowski structure, which is observable in principle for $m_g>0$. Instead, in GLET this background splits into absolute space and absolute time. This split is unobservable. Nonetheless, there remain some other differences as well.

\subsection{The choice of the signs of the cosmological terms}

First, the derivation of GLET does not lead to the restrictions for the sign of $\Xi,\Upsilon,\Lambda$.  Instead, the signs of the parameters $\Xi,\Upsilon,\Lambda$ in RTG are fixed. This gives more freedom of choice (but less predictive power) for GLET. In particular, the signs of the cosmological parameters have to be established by observation in GLET, while they are predicted in RTG.

Now, the choice of the sign of $\Upsilon$ as $\Upsilon > 0$ has a theory-independent natural justification. This choice prevents the two important GR singularities: The big bang becomes a big bounce (see app. \ref{homogeneousUniverse}), and the gravitational collapse stops before horizon formation. This choice is especially remarkable because, from a naive point of view, which ignores the important qualitative differences between the coordinates $\x^\a(x)$ and scalar fields $U^\a(x)$, the choice of the sign $\Upsilon > 0$ would be the wrong one. But this ``wrong'' sign, instead of leading to more instabilities, even prevents the formation of singularities.

For the cosmological term $\Lambda$, the situation is in one sense similar: The RTG choice of $\Lambda$ leads to more stability. The global evolution of the universe becomes, in this case, oscillating \cite{LogunovBB}. Unfortunately, this preferable theoretical choice seems to be in contradiction with observation: The universe seems to expand in an accelerating way, which may be nicely explained by a positive cosmological constant. Thus, in this case observation seems to prefer GLET, where the choice of the sign of $\Lambda$ is open. Instead, RTG requires some other explanation (some sort of ``dark energy'' or ``quintessence'') to fit observation, which is not necessary in GLET.

Last but not least, there is a nice general theoretical argument in favour of the signs used in RTG, which can be as well applied in GLET: The existence of a constant (vacuum) solution $g_{\mu\nu}(x)=\eta_{\mu\nu}$. This choice would require $\Xi>0$.

\subsection{The causality conditions}

Second, in above theories we have additional restrictions related with the notion of causality -- causality conditions.  In GLET, causality is related with the Newtonian background. The direction of causality is defined by the preferred time $\t(x)$. To be compatible with the usual interpretation of causality in metric theories of gravity, this preferred time $\t(x)$ should be a time-like function. In terms of the ether variables, this is equivalent to the condition $\rho>0$.

In RTG, causality is defined by the Minkowski background: The light cone of the physical metric $g_{\mu\nu}$ should be inside the light cone of the background metric $\eta_{\mu\nu}$. The RTG causality condition is, therefore, much more restrictive. Interesting solutions of GLET which violate the RTG causality condition and are, therefore, not acceptable as solutions of RTG, exist. Especially we have oscillating homogeneous universe solutions without matter (see appendix \ref{homogeneousUniverse}), while only the trivial solution $g_{\mu\nu}(x)=\eta_{\mu\nu}$ is compatible with the RTG causality condition.

A consequence of the RTG causality principle is that the parameter $\Xi$ has to be sufficiently small, thus, the background lightcone sufficiently wide, so that the universe remains inside the background lightcone all time. This makes the effects of $\Xi$ effectively unobservable (cf. \cite{Logunov1}).

These restrictions do not hold in GLET. The parameter $\Xi$ may be, therefore, sufficiently large to become observable.

\subsection{Summary}

Last but not least, there is a large amount of metaphysical differences. The motivation for above theories is completely different. GLET has been largely motivated by the problem of compatibility of a preferred frame with relativistic gravity, as well as the existence of a condensed matter interpretation (see section \ref{motivation}). These questions are of no interest from point of view of RTG, which is a relativistic field theory. These questions have been considered as well in app.  \ref{appUnobservable}.

We conclude that, while our theory of gravity is very close to RTG (comparable to the similarity between the Lorentz ether interpretation and the Minkowski spacetime interpretation of special relativity), these theories should be, nonetheless, considered as different physical theories. This is a consequence of two important physical differences:

\begin{itemize}
\item The different causality conditions of above theories;
\item The fact that the signs of the cosmological constants $\Xi,\Upsilon,\Lambda$ are not fixed by the derivation of GLET.
\end{itemize}

But even the other, metaphysical differences are important from physical point of view, because they lead to completely different programs for the further development of the two theories, in particular for their quantization: GLET suggests an atomic ether, which provides an explicit regularization in space but not time, and, therefore, breaking relativistic symmetry. Instead, RTG is a relativistic field theory and faces the problems of non-renormalizability of gravity.

\section{Quantization of Gravity Based on Ether Theory}
\label{appQuantization}

One of the most important problems of fundamental physics is quantization of gravity. Here, an ether theory suggests an alternative quantization concept which at least solves some of the most serious problems of GR quantization. Indeed, we already know one simple way to quantize classical condensed matter theories in a Newtonian framework: It would be sufficient to use classical canonical Schr\"{o}dinger quantization of some discrete (atomic) model of these field theories.

Such a discrete model, of course, has to incorporate also the matter fields, thus, in case of our actual universe, some discrete model for the standard model of particle physics (SM) or some more fundamental theory replacing it. The proposal of the author for such a discrete model has been given in \cite{clm}. This model already contains a quantization scheme for the fermionic part of the SM.

In this approach to the quantization of gravity as well as that of gauge fields, we do not need a separate ``quantization procedure'' for these fields: They appear only as effective fields of the more fundamental theory. Thus, if the degrees of freedom of the more fundamental theory have been quantized, everything has been quantized. Instead of a ``quantization'' of gravity and gauge fields, one has to derive the continuous field theory limit of the fundamental quantum theory, which will be some quantum field theory. 

The ether interpretation of the gravitational field in terms of density, velocity, and pressure tensor, which is independent of the particular model for the material properties (or matter fields) suggests that some steps of this derivation will be independent of these models too.

In particular, the description in terms of density, velocity, and pressure is based on Euler (local) variables. What has to be quantized are the movements of the atomic constituents of the ether, which are described by the Lagrange (material, comoving) variables.  Thus, one has to switch from Euler variables to Lagrange variables. It is known that a Lagrange formalism in Euler variables gives a Lagrange formalism in Lagrange variables and that this transformation gives a canonical transformation of the related Hamilton formalisms \cite{Broer}.

This essentially changes the situations with the constraints. The condensed matter is no longer described by $\rho(\x,\t)$, $v^i(\x,\t)$, but by $\x^i(\x^i_0,\t)$.  The continuity equation disappears. Instead, the Euler equation becomes a second order equation. Thus, instead of some first order equations (which become constraints in the Hamilton formalism) we obtain second order equations, which are much less problematic. In other words, the ether automatically solves the problems related with the first order conditions $\pd_\a(g^{\a\b}\sqrt{-g})$ which appear in GR in harmonic gauge.

A second general point is that, whatever the atomic model, it provides an explicit regularization of the continuous field theory. Thus, the ether automatically gets rid of ultraviolet infinities. Thus, the non-renormalizability of GR is not problematic at all in an ether approach, because we have a physical regularization defined by the atomic ether model. 

Last but not least, a third general property of an ether approach to the quantization of gravity is the different handling of symmetry. The fundamental ether model is, clearly, not supposed to show relativistic symmetry on the fundamental level. Relativistic symmetry is only a large distance emergent symmetry. A guidance how to derive this symmetry in the continuous limit is given by our derivation of the general Lagrangian: What we need to derive it is some Lagrange formalism for the discrete ether model which allows to derive a continous Lagrange formalism having the properties described by our axioms.

\section{Comparison with Canonical Quantization}

It is worth to note that this quantization program is completely different from the program of canonical quantization of general relativity: Space and time remain a classical background and are not quantized.  We have a different number of degrees of freedom: States which are different but indistinguishable for internal observers are nonetheless handled as different states.  That means, in related path integral formulations their probabilities have to be added.  No ghost fields have to be introduced.  There is also no holography principle: the number of degrees of freedom of the quantum theory is, for almost homogeneous $\rho$, proportional to the volume.

The problems which have to be solved in this quantization program seem to be not very hard in comparison with the problems of canonical GR quantization.  This suggests, on the other hand, that quantization of gravity following this scheme will not lead to non-trivial restrictions of the parameter space of the standard model: There will be a lot of quantum theories of gravity with very different material models.

That a preferred frame is a way to solve the ``problem of time'' in quantum gravity is well-known: ``... in quantum gravity, one response to the problem of time is to `blame' it on general relativity's allowing arbitrary foliations of spacetime; and then to postulate a preferred frame of spacetime with respect to which quantum theory should be written.''  \cite{Butterfield}.  This way to solve the problem is rejected not for physical reasons, but because of deliberate metaphysical preference for the standard general-relativistic spacetime interpretation: ``most general relativists feel this response is too radical to countenance: they regard foliation-independence as an undeniable insight of relativity.''  \cite{Butterfield}.  These feelings are easy to understand. At ``the root of most of the conceptual problems of quantum gravity'' is the idea that ``a theory of quantum gravity must have something to say about the quantum nature of space and time'' \cite{Butterfield}.  The introduction of a Newtonian background ``solves'' them in a very trivial, uninteresting way.  It does not tell anything about quantum nature of space and time, because space and time do not have any quantum nature in this theory -- they have the same classical nature of a ``stage'' as in non-relativistic Schr\"odinger theory.  The hopes to find something new, very interesting and fundamental about space and time would be dashed.  But nature is not obliged to fulfill such metaphysical hopes.

\section{Compatibility with Realism and Hidden Variables}
\label{appRealism}

An important advantage of a theory with preferred frame is related with the violation of Bell's inequality \cite{Bell}.  It is widely accepted that experiments like Aspect's \cite{Aspect} show the violation of Bell's inequality and, therefore, falsify Einstein-local realistic hidden variable theories.  Usually this is interpreted as a decisive argument against hidden variable theories and the EPR criterion of reality.  But it can as well turned into an argument against Einstein locality.  Indeed, if we take classical realism (the EPR criterion \cite{EPR}) as an {\em axiom\/}, the violation of Bell's inequality simply \emph{proves the existence} of superluminal causal influences.  Such influences are compatible with a theory with preferred frame and classical causality, but not with Einstein causality.

This has been mentioned by Bell \cite{Bell1}: ``the cheapest resolution is something like going back to relativity as it was before Einstein, when people like Lorentz and Poincare thought that there was an aether --- a preferred frame of reference --- but that our measuring instruments were distorted by motion in such a way that we could no detect motion through the aether. Now, in that way you can imagine that there is a preferred frame of reference, and in this preferred frame of reference things go faster than light.''

Note that at the time of the EPR discussion the situation was quite different.  People have accepted the rejection of classical realism under the assumtion that realistic hidden variable theories for quantum theory do not exist, and that this is a theorem proven by von Neumann.  Today we know not only that EPR-realistic, even deterministic hidden variable theories exist -- we have de Broglie-Bohm pilot wave theory \cite{Bohm}, \cite{BohmHiley} as an explicit example. Thus, quantum theory alone does not give any argument against classical realism, as was believed at that time.  Therefore, the only argument against classical realism is its incompatibility with Einstein causality --- thus, the reason why we mention it here as support for our ether theory.

We use here the existence of pilot wave theory as an argument for compatibility of classical realism with observation. Therefore, some remarks about objections against pilot wave theories are useful.  First, the main objection against pilot wave theories is that they need a preferred frame.  This is already a decisive argument for many scientists. But it cannot count here because it is the reason why we mention it as support for ether theory.  But there have been other objections: Problems with spin, fermions, and relativistic particles are often mentioned.  But these problems have been solved in modern presentations of pilot wave theories \cite{BohmHiley} \cite{Duerr}.  It seems useful to mention here that the pilot wave concept is a quite general concept which works on quite arbitrary configuration spaces.  Thus, it does not depend on a particle picture.  Once a quantum theory starts with fields, it's natural pilot wave interpretation is one with a field theory.

Thus, we can use BM here as support for our ether theory.  Reversely, our ether theory defines a program for Bohmian gravity.  Indeed, following the quantization program described in app. \ref{appQuantization} we end with canonical quantization of a discrete theory, which fits into the domain of applicability of the classical scheme of Bohmian mechanics.

\section{Analog Models of General Relativity}
\label{appCondensedMatter}

In this paper, we propose an ether theory of gravity, starting from condensed-matter-like axioms and deriving a metric theory of gravity with GR limit.  But this connection between condensed matter axioms and a metric theory with GR limit may be, possibly, applied also in another direction: Search for classical condensed matter theories which fit into this set of axioms as analog models of general relativity. In this appendix we consider the possibility of such applications.

\subsection{Comparison With Analog Gravity Research}

The observation that curved Lorentz metrics appear in condensed matter theories is currently attracting considerable attention.  Though it is not a priori expected that all features of Einstein gravity can successfully be carried over to the condensed matter realm, interest has turned to investigating the possibility of simulating aspects of general relativity in analog models.  Domains of research in ``over a hundred articles devoted to one or another aspect of analog gravity and effective metric techniques'' have been dielectric media, acoustics in flowing fluids, phase perturbations in Bose-Einstein condensates, slow light, quasi-particles in superfluids, nonlinear electrodynamics, linear electrodynamics, Scharnhorst effect, thermal vacuum, ``solid state'' black holes and astrophysical fluid flows \cite{Barcelo}.  Asking whether something more fundamental is going on, Barcelo et al \cite{Barcelo} have found that linearization of a classical field theory gives, for a single scalar field, a unique effective Lorentz metric.  As observed by Sakharov \cite{Sakharov}, quantization of the linearized excitations around this background gives a term proportional to the Einstein-Hilbert action in the one-loop effective action.  But this approach has serious limitations: It is not clear how to obtain the Einstein equivalence principle for different types of excitations which, in general, ``see'' different effective metrics.  The Einstein-Hilbert action appears only in the quantum domain.  The most serious question is how to suppress the non-covariant terms.

Now, the approach presented here may be used to solve these problems. If we are able to find some real condensed matter which fits into our set of axioms, this matter defines an ideal analog model of general relativity.  Thus, we have reduced the search for analog models for general relativity to a completely different question -- the search for Lagrange formalisms for condensed matter theories which fit into our set of axioms.  While we have not been able to find such Lagrange formalisms yet, there are some points which seem worth to be mentioned:

\begin{itemize}
\item  Our metric has been defined independent of any particular linearization of waves.  It's definition is closely related to the energy-momentum tensor:

\begin{equation}
T^\gamma_\alpha = -(4\pi G)^{-1}\gamma_{\alpha\beta} g^{\beta\gamma}\sqrt{-g}
\end{equation}

It is in no way claimed that some or all excitations which may be obtained by various linearizations follow this metric.  Such a claim would be false even in the world of standard general relativity. Indeed, we know particular excitations which have different speeds -- like light on the background of a medium.  The EEP allows to distinguish ``fundamental'' excitations (which use its basic metric) and ``less fundamental'' excitations (like light in a medium).  But starting with arbitrary waves in a medium described phenomenologically or obtained by some linearization of approximate equations, without knowing the EEP-related metric, we have no base to make this distinction.

Once we do not define the metric as an effective metric of some excitations, it is not a problem for our approach if different excitations follow different effective metrics.  The metric is uniquely defined by the energy-momentum tensor, which is closely related to translational symmetry.  Therefore it is not strange that this metric is the metric related with the EEP.

\item The general-relativistic terms appear already in the classical domain.  Their origin is not quantization, but the general restrictions for non-covariant terms which follow from our axioms. Whatever additional terms we add to the Lagrangian -- if the modified theory does not violate the axioms and does not renormalize $\gamma_{\alpha\beta}$, the difference between old and new Lagrangian should be covariant.

\item The non-covariant terms in our approach are uniquely defined and easy to control: they do not depend on matter fields, and they do not depend on derivatives of $g_{\a\b}$.  

\end{itemize}

It remains to find interesting condensed matter theories which fits into our axioms. This problem has to be left to future research. 

\subsection{Condensed Matter Interpretation of Curvature}

The physical meaning of the general-relativistic Lagrangian can be understood as related with inner stress.  If at a given moment of time all spatial components of the curvature tensor vanish, then there exists a spatial diffeomorphism which transforms the metric into the Euclidean metric.  Such diffeomorphisms are known as transformations into stress-free reference states in elasticity theory.  If no such transformation exists, we have a material with inner stress.  This understanding is in agreement with a proposal by Malyshev \cite{Malyshev} to use of the 3D Euclidean Hilbert-Einstein Lagrangian for the description of stress in the presence of dislocations.

The physical meaning of the components which involve time direction is less clear.  It seems clear that they describe modifications of inner stresses, but this has to be better understood.  Nonetheless, our approach suggests a generalization of Malyshev's proposal -- to use the full Hilbert-Einstein Lagrangian instead of its 3D part only.

\subsection{Condensed Matter Interpretation of Gauge Fields and Fermions}
\label{appGauge}

Of course, search for analog models requires a better understanding of the matter part of the theory.  It would be helpful to have a similar understanding of gauge and fermion fields. It seems natural to interpret the analog of the harmonic condition, the Lorenz gauge condition

\begin{equation}\label{Lorenz}
\partial_\mu A^\mu = 0,
\end{equation}

as a conservation law.  Then, gauge symmetry will be no longer a fundamental symmetry.  Instead, it should be derived in a similar way as the EEP, describing the inability of internal observers to distinguish states which are different in reality.

The example of superfluid $^3He-A$ seems to be of special interest: ``the low-energy degrees of freedom in $^3He-A$ do really consist of chiral fermions, gauge fields and gravity'' \cite{Volovik}.

Another proposal for an ether interpretation of the standard model fermions and gauge fields has been made by the author \cite{clm}.

\section{Cosmological Predictions}
\label{appPredictions}

It is not the purpose of this paper to consider the various experimental differences between GR and our theory in detail. Nonetheless, we want to give here a short description of some differences to support the claim that such differences exist and lead to observable effects.  It is clear that the predictions have to be worked out in detail.

In general, because the Einstein equations appear as the limit $\Xi,\Upsilon\to 0$, the overwhelming experimental evidence in favor of the EEP and the Einstein equations (Solar system tests, binary pulsars and many others) as described by Will \cite{Will} can give only upper bounds for the parameters $\Xi, \Upsilon$.  It is not the purpose of this paper to obtain numerical values for these bounds.

Some qualitative differences have been already mentioned: non-trivial topologies and closed causal loops are forbidden.  There should exist a global time-like harmonic coordinate.  Additional predictions require reasonable guesses about the preferred coordinates.  But this is easy.  For the global universe, we obtain such natural guesses based on obvious symmetry considerations.  For local situations (gravitational collapse) we have the requirement that the preferred coordinates have to fit the global solution and need reasonable initial values.  Together with the harmonic equation, this is enough to define them.

Note that all the differences between GLET and GR considered below depend on the assumptions about the special geometric nature of the preferred coordinates.  Especially the GLET solutions have the boundary conditions appropriate for coordinates, but not for scalar matter fields.  Therefore, the difference between GR with scalar fields and GLET is not only a purely theoretical one, it leads to really observable differences.

\subsection{The Homogeneous Universe}\label{homogeneousUniverse}

The background-dependent terms of GLET lead to interesting observable effects.  Let's consider at first the homogeneous universe solutions of the theory.  Because of the Newtonian background frame, only a flat universe may be homogeneous.  An appropriate harmonic ansatz for the flat homogeneous universe is

\begin{equation} ds^2 = a^6(\t) d\t^2 - \beta^4 a^2(\t)(dx^2+dy^2+dz^2). \end{equation}

For proper time $d\tau = a^3 d\t$ this gives the usual FRW-ansatz for the flat universe with some scaling factor $\beta>0$:

\begin{equation} ds^2 = d\tau^2 - \beta^4a^2(\tau)(dx^2+dy^2+dz^2). \end{equation}

This leads to ($p=k\varepsilon$, $8\pi G = c = 1$):

\begin{eqnarray}\label{energyEquation}
3\left(\frac{\dot{a}}{a}\right)^2  &=&
        - \frac{\Upsilon}{a^6} + 3 \frac{\Xi}{\beta^4a^2} + \Lambda + \varepsilon\\
2\frac{\ddot{a}}{a} + \left(\frac{\dot{a}}{a}\right)^2 &=&
        + \frac{\Upsilon}{a^6} +   \frac{\Xi}{\beta^4a^2} + \Lambda - k \varepsilon
\end{eqnarray}

The $\Upsilon$-term influences only the early universe, its influence on later universe may be ignored.  But, if we assume $\Upsilon >0$, the qualitative behavior of the early universe changes in a remarkable way.  We obtain a lower bound $a_0$ for $a(\tau)$ defined by

\begin{equation} 
\frac{\Upsilon}{a_0^6} = 3 \frac{\Xi}{\beta^4a_0^2} + \Lambda + \varepsilon  
\end{equation}

The solution becomes symmetrical in time -- a big bounce.  Note that this solves two problems of cosmology which are solved in standard cosmology by inflation theory: the flatness problem and the cosmological horizon problem.\footnote{About these problem, see for example Primack \cite{Primack}: First, ``the angular size today of the causally connected regions at recombination ($p^+ + e^-\to H$) is only $\Delta\theta\sim 3^o$.  Yet the fluctuation in the temperature of the cosmic background radiation from different regions is very small: $\Delta T/T\sim 10^{-5}$. How could regions far out of causal contact have come to temperatures that are so precisely equal?  This is the `horizon problem'.''  (p.56) Even more serious seems the following problem: In the standard hot big bang picture, ``the matter that comprises a typical galaxy, for example, first came into causal contact about a year after the big bang.  It is hard to see how galaxy-size fluctuations could have formed after that, but even harder to see how they could have formed earlier'' (p.8).}  The question what the theory predicts about the distribution of fluctuations of the CMBR radiation requires further research: It depends on reasonable initial conditions before the bounce.  The bounce itself is too fast to allow the establishment of a global equilibrium, thus, the equilibrium should exist already before the bounce. 

Instead, the $\Xi$-term gives a dark energy term. For $\Xi >0$ it behaves like homogeneously distributed dark matter with $p=-(1/3)\varepsilon$.  The additional term has the same influence on $a(\tau)$ as the classical curvature term.  On the other hand, the universe should be flat. \footnote{This makes the comparison with observation quite easy: as long as $a(\tau)$ is considered, the prediction of a $\Xi$CDM theory with $\Lambda=0, \Xi>0$ is the same as for OpenCDM.  Therefore, the SNeIa data which suggest an accelerating universe are as problematic for $\Xi$CDM as for OpenCDM. The 2000 review of particle physics \cite{RPP2000}, 17.4, summarizes: ``the indication of $\Omega_\Lambda\ne 0$ from the SNeIa Hubble diagram is very interesting and important, but on its own the conclusion is susceptible to small systematic effects.  On the other hand small scale CBR anisotropy confirming a nearly flat universe ... strongly suggests the presence of $\Lambda$ or other exotic (highly negative pressure) form of dark mass-energy.''  The fit with the farthest SN 1997ff can be seen in in figure 11 of \cite{Riess} -- the figure for $\Omega_M=0.35,\Omega_\Xi=0.65$ is the same as for the $\Omega_M=0.35,\Omega_\Lambda=0$ open universe given in this figure. Thus, while not favored by the SNeIa results, $\Xi$CDM may be nonetheless a viable dark energy candidate, better than OpenCDM because it requires a flat universe.}

The case of a universe without matter $\varepsilon(\tau)=0$ is interesting for comparison with RTG \cite{Logunov1}.  We obtain a vacuum solution with $a^4(\tau)\Xi=\beta^4\Upsilon$, $a(\tau)^6\Lambda=2\Upsilon$. The RTG causality condition gives $a^4(\tau)\Xi\le\beta^4\Upsilon$, so that this is also a valid RTG solution. But we also obtain oscillating solutions for smaller values of $\beta$, and these oscillating solutions already violate the RTG causality condition. 

\subsection{Gravitational Collapse}\label{collapse}

Another domain where $\Upsilon>0$ leads to in principle observable differences to GR is the gravitational collapse.  Let's consider for this purpose a stable spherically symmetric metric in harmonic coordinates.  For a function $m(r), 0<m<r, r=\sqrt{\sum (\x^i)^2}$, the metric

\begin{equation} ds^2 = (1-\frac{m \partial m/\partial r }{ r})
                (\frac{r-m}{ r+m}d\t^2-\frac{r+m}{ r-m}dr^2)
        - (r+m)^2 d\Omega^2 \end{equation}

is harmonic in $\x^i$ and $\t$.  For constant m this is the Schwarzschild solution in harmonic coordinates.  For time-dependent $m(r,t)$ the $\x^i$ remain harmonic but $t$ not.  $T$ has to be computed by solving the harmonic equation for Minkowski initial values before the collapse.

Now, a non-trivial term $\Upsilon>0$ becomes important near the horizon.  Indeed, let's consider as a toy ansatz for the inner part of a star the distribution $m(r)=(1-\Delta)r$.  This gives

\begin{eqnarray*}
ds^2 &=& \Delta^2dt^2 - (2-\Delta)^2(dr^2+r^2d\Omega^2) \\
0    &=& -\Upsilon \Delta^{-2} +3\Xi(2-\Delta)^{-2}+\Lambda+\varepsilon\\
0    &=& +\Upsilon \Delta^{-2} + \Xi(2-\Delta)^{-2}+\Lambda-p
\end{eqnarray*}

We can ignore the cosmological terms $\Xi, \Lambda$ in comparison to the matter terms $\varepsilon, p$, but for very small $\Delta$ even a very small $\Upsilon>0$ becomes important.  Our ansatz gives a stable ``frozen star'' solution for ultra-relativistic $p=\varepsilon$ and $\Xi=\Lambda=0$.  The surface time dilation is $\Delta^{-1}=\sqrt{\varepsilon/\Upsilon} \sim M^{-1}$ for mass M.  In general, it seems obvious that the term $\Upsilon>0$ prevents $g^{00}\sqrt{-g}$ from becoming infinite in harmonic coordinates. Therefore, the gravitational collapse stops and leads to some stable frozen star with a size very close to the Schwarzschild radius.

In principle, this leads to observable effects.  The surface of the objects considered as black holes in GR should be visible but extremely redshifted: Radiation emitted with frequence $\nu$ would be visible as having the frequence $(1+z)^{-1}\nu$ for the surface redshift $z$, which may be extremely large. The question is, therefore, about the temperature of the surface of the star. If it would be correspondigly blueshifted, the surface would be visible as a usual star. Now, infalling matter has, if it arrives at the surface, in terms of the suface system of coordinates a corresponding blueshift. This is a simple consequence of energy conservation. The question is, therefore, about the dependence of the full energy of the star from it's temperature $U=U(T)$. We should know this dependence for extremely large (extremely blueshifted) temperatures $T$ to predict the temperature $(1+z)^{-1}T$ of the surface as visible from outside. Moreover, we have to rely on the reliability of the Einstein equivalence principle to connect the predictions for extremal time dilations with those without time dilation.

An ideal point-like elastic infalling body would bounce back from the surface of a frozen star. This has been found for RTG in \cite{LogunovBH} and remains correct for $\Upsilon>0$  in our theory. If this leads to easily observable effects is another question. Real infalling matter will clearly not remember an ideal elastic body in such an extremal situation. For extremeal redshifts, it would have to preserve ideal elasticity even for extremal energies. Nonetheless, one can hope that at least some part of the infalling matter will bounce back, giving some visible evidence for the existence of a surface. If this happens is a question which has to be left to future research.

Instead, the choice $\Upsilon<0$ would lead to horizon formation and to a singularity inside the black hole, similar to GR. This would be, at the current moment, clearly a viable alternative, but certainly not a beautiful one. 

\subsection{Ether Sound Waves and Gravitational Waves}

In fluids described by continuity and Euler equations there are sound waves.  This seems to be in contradiction with the situation in general relativity.  They cannot correspond to gravitational waves as predicted by general relativity, which are quadrupole waves.

Now, in the Lorentz ether these waves are gravity waves.  Because general-relativistic symmetry is broken by the additional background-dependent terms, the gravitational field has more degrees of freedom in the Lorentz ether.  In some sense, these additional degrees of freedom are the four fields which appear in GRCF discussed in app. \ref{appGRCF}.  Now, these additional degrees of freedom interact with the other degrees of freedom only via the background-dependent term. That means, they interact with matter only as dark matter.  Second, in the limit $\Xi,\Upsilon\to 0$ they no longer interact with usual matter at all.  Thus, in this limit they become unobservable for internal observers.  That's why in this limit they may be removed from considerations as hidden variables.

On the other hand, for non-zero $\Xi,\Upsilon$ they nonetheless interact with matter.  And this interaction leads, in principle, to observable effects.  Nonetheless, we have not considered such effects, because we believe that the restrictions which follow from cosmological considerations are much stronger.  This belief is qualitatively justified by our considerations about the GR limit: We have argued that the GR limit is relevant for small distance high frequency effects, and if the total variation of the $g_{\mu\nu}$ remains small.  This is exactly the situation for gravitational waves.

On the other hand, there may be other factors which override these simple qualitative considerations.  For example the ability of gravity waves to go over large distances which leads to integration of these small effects over large distances, or for measurements of energy loss as done for binary pulsars, which allows integration over many years. For the range of gravity waves in theories with massive graviton such considerations have been made in \cite{Logunov3}, and their results agree with our expectation that the upper bounds coming from cosmological considerations are nonetheless much stronger.

\end{appendix}

\section{Acknowledgements} I would like to thank A.A. Logunov, V.A. Petrov and A.B. Shabat for interesting and helpful conversations, and W.A. Rodrigues Jr. for identifying some weak places in the article.

\end{document}